\documentclass[3p]{elsarticle}
\usepackage{lineno,hyperref}
\usepackage{amsmath}
\usepackage{color}
\usepackage{mathtools}
\usepackage[normalem]{ulem}
\usepackage{rotating}
\usepackage{makecell}
\usepackage{threeparttable, tablefootnote}
\usepackage{subcaption}
\newcommand{\Beta}{\mathrm{B}}

\journal{arXiv}
\newcommand\myFigureWidth{0.42}

\begin{document}
\begin{frontmatter}

\title{Modeling Response Time Distributions with Generalized Beta Prime}

 \author[mymainaddress]{M. Dashti Moghaddam}
 \author[mymainaddress]{Jiong Liu}
 \author[mysecondaryaddress]{John G. Holden\fnref{hisfootnote}}
 \author[mymainaddress]{R. A. Serota\fnref{myfootnote}}
 \fntext[myfootnote]{serota@ucmail.uc.edu}
 \fntext[hisfootnote]{holdenjn@ucmail.uc.edu}

 \address[mymainaddress]{Department of Physics, University of Cincinnati, Cincinnati, Ohio 45221-0011}
 \address[mysecondaryaddress]{Department of Psychology, CAP Center for Cognition, Action, and Perception, University of Cincinnati, Cincinnati, Ohio 45221-0376}

\begin{abstract}
We use Generalized Beta Prime distribution, also known as GB2, for fitting response time distributions. This distribution, characterized by one scale and three shape parameters, is incredibly flexible in that it can mimic behavior of many other distributions. GB2 exhibits power-law behavior at both front and tail ends and is a steady-state distribution of a simple stochastic differential equation. We apply GB2 in contrast studies between two distinct groups -- in this case children with dyslexia and a control group -- and show that it provides superior fitting. We compare aggregate response time distributions of the two groups for scale and shape differences (including several scale-independent measures of variability, such as Hoover index), which may in turn reflect on cognitive dynamics differences. In this approach, response time distribution of an individual can be considered as a random variate of that individual's group distribution. \\ \\ \\
\end{abstract}


\end{frontmatter}

\section{Introduction}

Response time (RT) is the time interval between the presentation of a stimulus to a subject and subject's response. Distributions of RT data, collected during cognitive tasks, can be studied to glean insights into underlying cognitive dynamics. One particular application is a comparison of distributions, in the form of probability density functions (PDF), of two distinct groups of individuals: for instance, a group of dyslexic children with a control group \cite{holden2014dyslexic} or a group of ADHD children receiving medication with those receiving placebo \cite{epstein2011effects}. There are three key elements in this process: \textit{aggregation, fitting and modeling}.

\subsection{Aggregation \label{aggregation}}
There is a range of performances within each group, which can be quite wide. For instance, dyslexia can vary from mild to severe and the effect of an ADHD medication can vary from significant to negligible. Aggregation of RT data of all individuals in the group serves two important purposes \cite{liu2016probability,liu2019modeling}. First, it allows to establish differences between distinct groups, such as existence of an \textit{overall} medication effect (or lack thereof). Second it helps to considerably improve statistics, which is particularly important for distributions with power-law (``fat") tails \cite{holden2002fractal}. If the PDF, including its parameters, of a group reflects group's cognitive dynamics, one can think of each individual in the group as a random variate generated from this PDF \cite{liu2016probability,liu2019modeling}. Of course, one cannot infer an individual subject outcome based on the group performance \cite{estes1956problem,estes2005risks}. In that case, RT distribution of an individual subject needs analyzed, for instance to determine the degree of dyslexia of (or the effect of medication on) that particular subject. 

\subsection{Fitting \label{fitting}}

To this day, short-tailed distributions characterized by (modified) exponential decay, are used to fit RT distributions. Two prominent examples are exponentially modified Gaussian (exGaussian) and translated Weibull (3-parameter Weibull). Both are characterized by three parameters: location, scale and shape. As shown in Table \ref{KSArithControl} they perform considerably worse than distributions with fat tails. Detailed discussion of other fitting candidates was conducted in \cite{liu2016probability,liu2019modeling}. In particular, thorough fitting was conducted using Generalized Inverse Gamma (GIGa) distribution and lognormal-Pareto "cocktail" (LNP) distribution \cite{holden2012self}. 

GIGa distribution has a location, a scale and two shape parameters. (From Table \ref{KSArithControl} it is clear that fitting with a 3-parameter GIGa, GIGa3, is almost as good since, in this case, location parameter of GIGa is always close to zero.) It is a steady-state distribution of a simple stochastic differential equation (SDE) (see below and \cite{hertzler2003stochastic,dashti2019implied,dashti2019generalized}) and is encountered in the context of a network model of economy \cite{ma2013distribution}. Its chief deficiency is that it decays as a (modified) exponent at front end. LNP matches lognormal distribution at front with a mixture of lognormal and Pareto so that the tail end is Pareto (power-law). LNP has one scale and three shape parameter. Its chief deficiency is that all its derivatives above first are discontinuous at the matching point and it does not appear as a result of a first-principles model.

Here, for the first time, we propose to use GB2 for fitting RT distributions \cite{mcdonald2008modelling}. It is given by
\begin{equation}
\label{gb2}
GB2(x; p, q, \alpha, \beta) = \frac{\alpha(1+(\frac{x}{\beta})^{\alpha})^{-p-q}(\frac{x}{\beta})^{-1+p\alpha}}{\beta \space \Beta(p,q)}
\end{equation}
where $B(p,q) = \frac{\Gamma(p)\Gamma(q)}{\Gamma(p+q)}$ is a Beta function, $\Gamma(\alpha)$ is a Gamma function, $p$, $q$ and $\alpha$ are shape parameters and $\beta$ is a scale parameter. Its limiting behaviors are given by 
\begin{equation}
GB2(x; p,q,\alpha,\beta) \propto (\frac{x}{\beta})^{-\alpha q -1}, \hspace{.25cm} x \gg \beta
\label{GB2largex}
\end{equation}
and
\begin{equation}
GB2(x; p,q,\alpha,\beta) \propto (\frac{x}{\beta})^{\alpha p -1}, \hspace{.25cm} x \ll \beta
\label{GB2smallx}
\end{equation}
We assume that $\alpha p>1$, that is that GB2 has a maximum (a bell shape). The condition for variance to exist is $\alpha q>2$. Clearly, for large  $\alpha p$ and $\alpha q$ GB2 can mimic lognormal and (modified) exponential behavior (formally, GIGa3 and GGa are, respectively, $p \rightarrow \infty$ and $q \rightarrow \infty $ limits of GB2 \cite{mcdonald1995generalization}). An important property of GB2 is that under a change of variable to its inverse, $x \to x^{-1}$ -- which converts front end to tail and vice versa -- it transforms as
\begin{equation}
GB2 (x; p, q, \alpha, \beta) \to GB2(x^{-1}; q, p, \alpha, \beta^{-1})
\label{GB2transform}
\end{equation}
which explains the $p+\frac{1}{\alpha} \leftrightarrow q$ symmetry of the averages calculated with GB2  \cite{dashti2019generalized}.

\subsection{Modeling \label{modeling}}

Consider a stochastic differential equation.
\begin{equation}
\mathrm{d}x = -\eta(x - \theta x^{1-\alpha})\mathrm{d}t + \sqrt{\kappa_2^2 x^2 + \kappa_\alpha^2 x^ {2-\alpha}}\mathrm{d}W_t
\label{GB2SDE}
\end{equation}
where $\mathrm{d}W_t$ is a Wiener process. Its steady-state distribution is  \cite{hertzler2003stochastic, dashti2019implied, dashti2019generalized} $GB2(x; p, q,\alpha, \beta)$ with
\begin{equation}
\label{betaGB2}
\beta=(\frac{\kappa_\alpha}{\kappa_2})^{2 / \alpha}
\end{equation}
\begin{equation}
\label{pGB2}
p=\frac{1}{\alpha}(-1+\alpha +\frac{2 \eta \theta}{\kappa_\alpha^2})
\end{equation}
and
\begin{equation}
\label{qGB2}
q=\frac{1}{\alpha}(1+\frac{2 \eta}{\kappa_2^2})
\end{equation}
The steady-state distribution of (\ref{GB2SDE}) is GIGa3 for $\kappa_{\alpha}=0$ and Generalized Gamma for $\kappa_2=0$. For $\alpha=1$ we have mean-reverting models which yield a Beta Prime steady-state distribution in general \cite{dashti2018combined} and Inverse Gamma and Gamma for $\kappa_1=0$ and $\kappa_2=0$ respectively. 

\clearpage
This is not to claim that eq. \ref{GB2SDE} necessarily describes a feature of cognitive dynamics. However, it belongs to the family of "birth and death" or "exchange" stochastic models \cite{holden2013change} that intend to describe natural and social phenomena. We strongly believe that such an underlying first-principles model should exist if a distribution is to be used for fitting purposes. Weibull, Pareto \cite{hertzler2003stochastic} and lognormal \cite{gualandi2018human} distributions, for instance, satisfy this requirement.

In this paper we apply methodology developed in \cite{liu2016probability} and \cite{liu2019modeling} to a study that compared RT performance in Flanker, Color Naming, Word Naming and Arithmetic tasks between children diagnosed with dyslexia and a control group of age-matched children (for a detailed description of tasks  see \cite{holden2014dyslexic}). Ordinarily, the deficiency in performance is associated with a slower RT. Initial question we wanted to address was whether the aggregate distributions are simply scaled versions of each other by the ratio of their mean RT \cite{holden2013cognitive}. We use a distribution-agnostic test by comparing zeroed-out distributions of log-transformed RT data. We find that for Flanker, where ratio of the mean control to dyslexic RT is the largest, rescaling cannot be rejected. For Color it can but not as strongly as for Word Naming and Arithmetic for which the ratios are the smallest.

Next, we wanted to quantify differences in aggregate distributions. To this end, we concentrated on their shape parameters and quantities associated with them. We fitted the distributions with GB2 and used bootstrap \cite{efron1994introduction} to obtain confidence intervals (CI) for the shape parameters and variability indices. Since the usual measures of variability, such as standard deviation (SD) are not ideal because they overweigh fat tails, we previously introduced half-width (HW) of the distribution as a measure of variability \cite{liu2016probability} and \cite{liu2019modeling}. However both SD and HW are proportional to the scale parameter. Mean-reduced SD \cite{epstein2011effects}, that is SD divided by the distribution mean, is scale-independent and so is the mean-reduced HW, rHW. We also use a quantity known in economics as the Hoover (Pietra, Schultz) inequality index \cite{mcdonald2008modelling}. Altogether, in addition to obtaining distribution and variability parameters and their CI, we find complete agreement between fitting results and simple scaling analysis.

This paper is organized as follows. In Section \ref{Rescaling} we conduct a test of whether the control and dyslexic distributions are scaled versions of each other. In Section \ref{GB2fit} we conduct thorough Maximum Likelihood Estimation (MLE) fitting of the distributions with GB2, using Kolmogorov-Smirnov (KS) statistics for goodness of fit, and use bootstrap to establish CI for important parameters, such as tail exponents and rHW and Hoover indices. The fitting is performed on the data with the un-physical small values of RT cut. We conclude with the Summary of the results. Appendix shows fits of the uncut data.

\section{Check for Rescaling \label{Rescaling}}

Here we conduct an agnostic -- that is not tied to fitting with a specific distribution -- check whether the aggregate distributions of control and dyslexic groups are just two versions of the same distribution scaled by the ratio of their mean values. The detailed description of methodology can be found in \cite{liu2016probability}. The results are summarized in Table \ref{ksLogTransformed}. Here $\mu_C$ and $\mu_D$ are mean values, measured in seconds -- of control and dyslexic group RT distributions respectively, $c_1 = \mu_C/\mu_D$, and $log(c_2)$ is the difference of the mean values of the distributions of the log-transformed control and dyslexic RT data -- theoretically $c_2$ and $c_1$ should be the same. KS statistics and p-values are obtained by comparing zeroed-out (moved to have zero mean) log-transformed data. Lower KS numbers correspond to a better match between distributions. 

\begin{figure*}[!htbp]
\centering
\begin{tabular}{cc}
\includegraphics[width = 0.49 \textwidth]{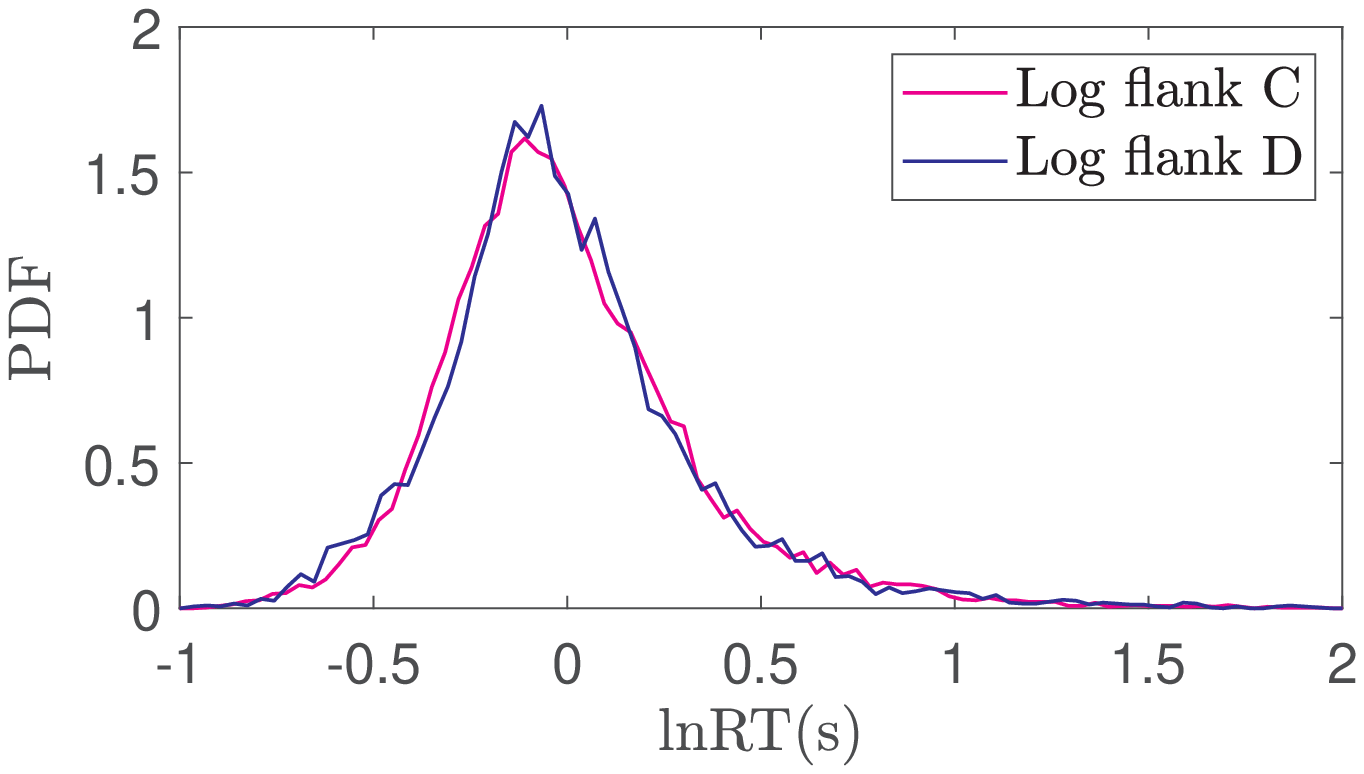} 
\includegraphics[width = 0.49 \textwidth]{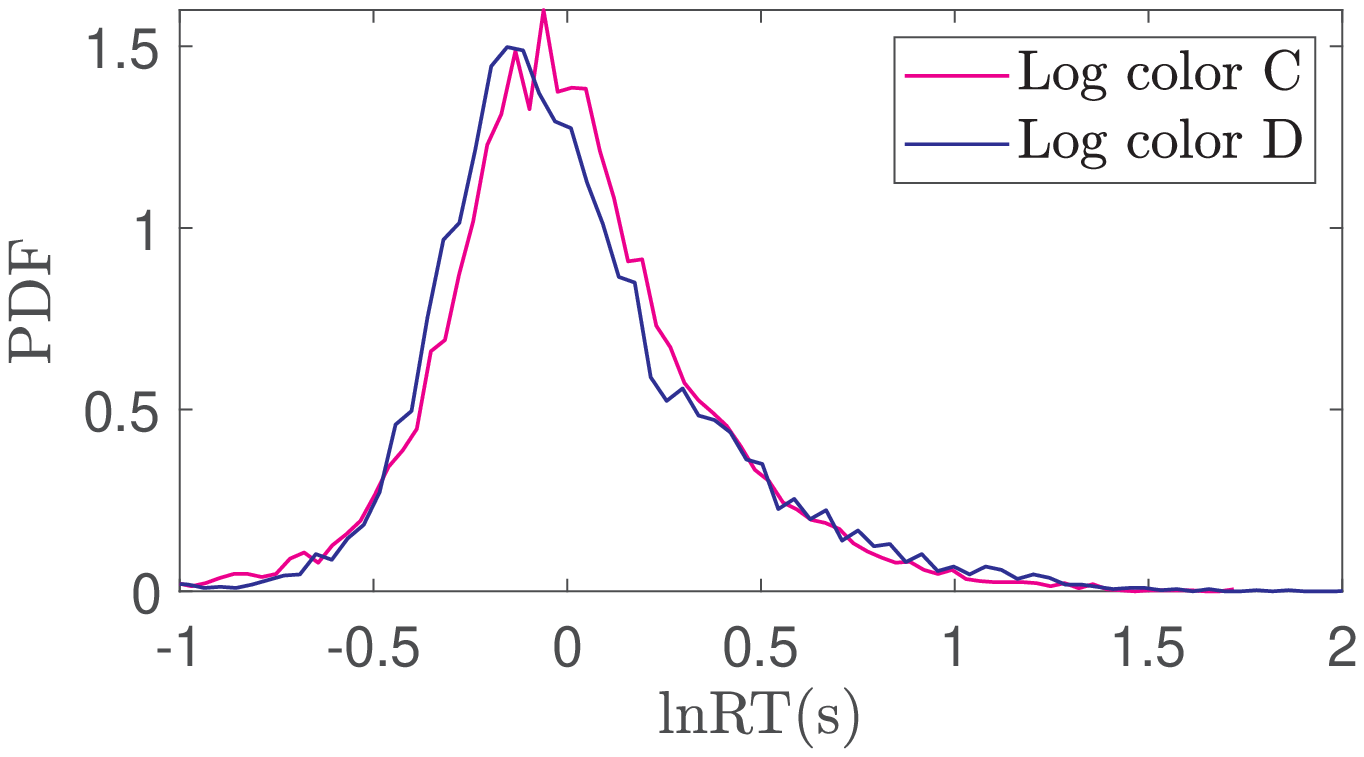}  \\ \\ \\ \\ 
\includegraphics[width = 0.49 \textwidth]{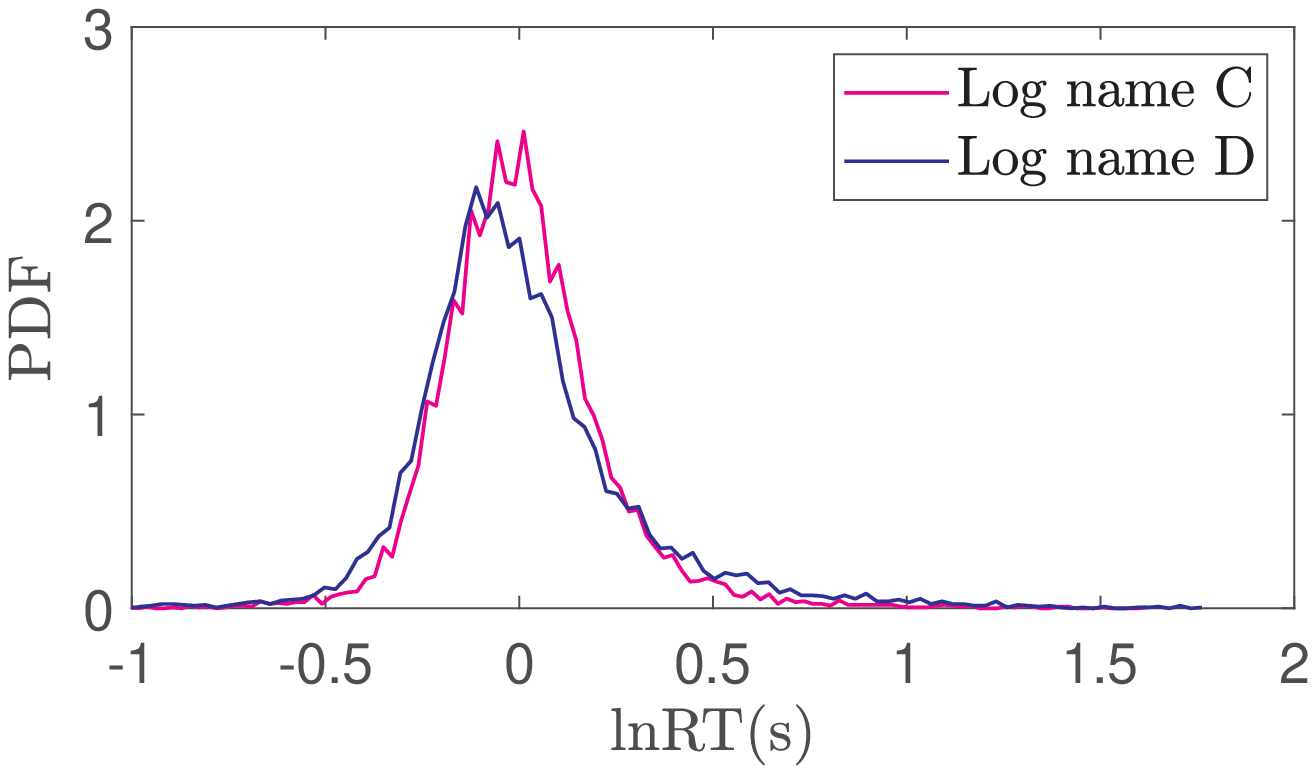} 
\includegraphics[width = 0.49 \textwidth]{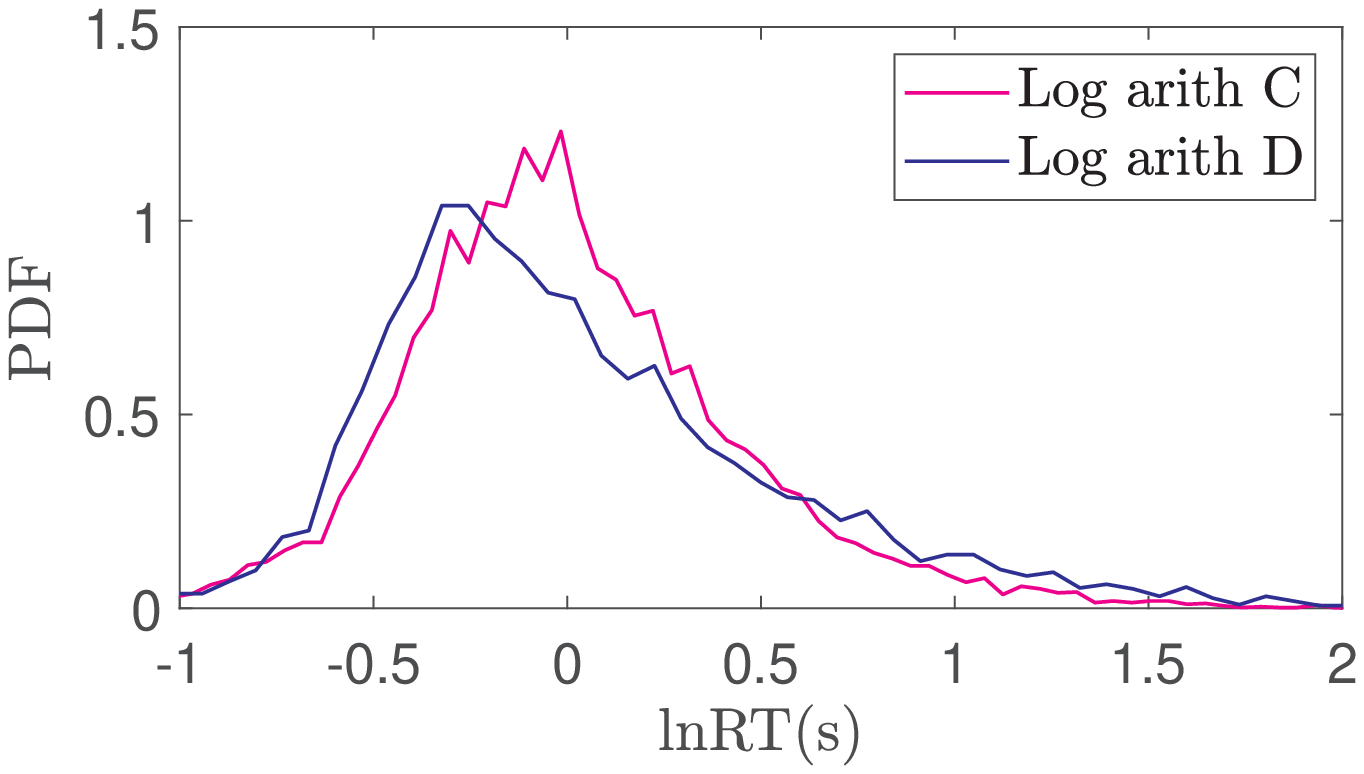} \\
\end{tabular}
\caption{Contour plots of zeroed-out (zero mean) distributions of log-transformed RT data of control and dyslexic groups.}
\label{loglog}
\end{figure*}

It is clear that for Flanker it cannot be ruled out that the control and dyslexic groups are scaled versions versions of each other, while the other three can -- with much greater certainty for Word Naming and Arithmetic tasks. Visually, this is illustrated in Fig. \ref{loglog}, which shows contour plots of zeroed-out log-transformed distributions. 

\begin{table}[!htbp]
\centering
\caption{KS test results of log-transformed data}
\label{ksLogTransformed}
\begin{tabular}{ccccccc} 
\hline
	& $\mu_C$ & $\mu_D$ & $c_1$ & $c_2$ & KS test statistics & KS test p-value \\
\hline
Flanker & 0.535 & 0.562 & 0.941 & 0.944 & 0.019 & 0.061 \\
\hline
Color Naming &0.707 & 0.828&0.854 & 0.865 & 0.045 &$5.458\times 10^{-8}$ \\
\hline
Word Naming & 0.554 & 0.694 & 0.798 & 0.814 & 0.071 &$7.772\times 10^{-20}$ \\
\hline
Arithmetic & 1.178 & 1.527 & 0.771 & 0.813 & 0.090 & $1.970\times 10^{-26}$ \\
\hline\\ \\
\end{tabular}
\end{table}

\clearpage
\section{Fitting with GB2 \label{GB2fit}}

\subsection{Maximum Likelihood Estimation Fitting and Kolmogorov-Smirnov Statistics}
Initially, we eliminated a number of un-physically small RT, which most likely were recorded in error. This was done by cutting in steps larger and larger number of low-time RT. At each step MLE fitting was performed and KS statistic obtained. The procedure continued until 6\% of the total was reached. The step with the lowest KS was identified and RT from all steps prior were cut. In Appendix we show the results of MLE fitting without cuts -- they produce consistently larger KS numbers (poorer fits).

\begin{figure*}[!htbp]
\centering
\begin{tabular}{cc}
\includegraphics[width = \myFigureWidth \textwidth]{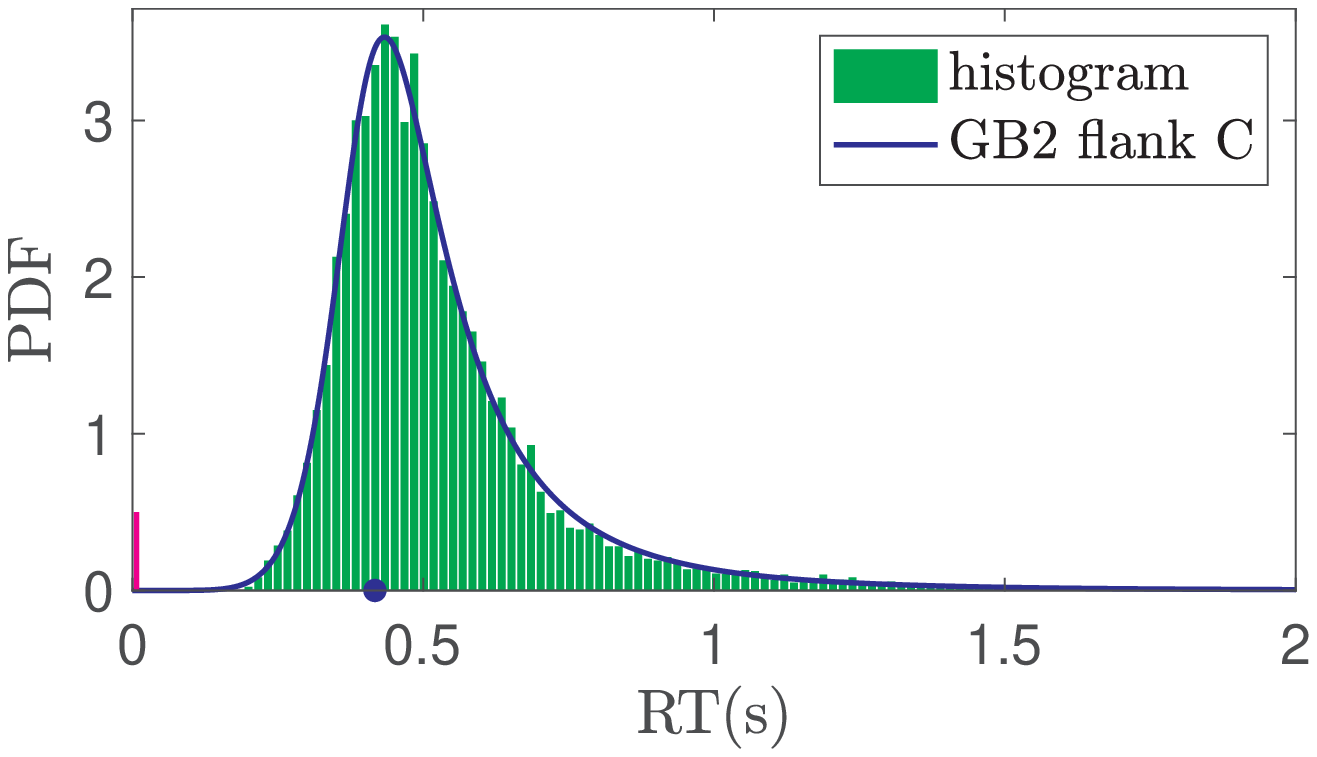}  \\ \\ \\ \\ \\
\includegraphics[width = \myFigureWidth \textwidth]{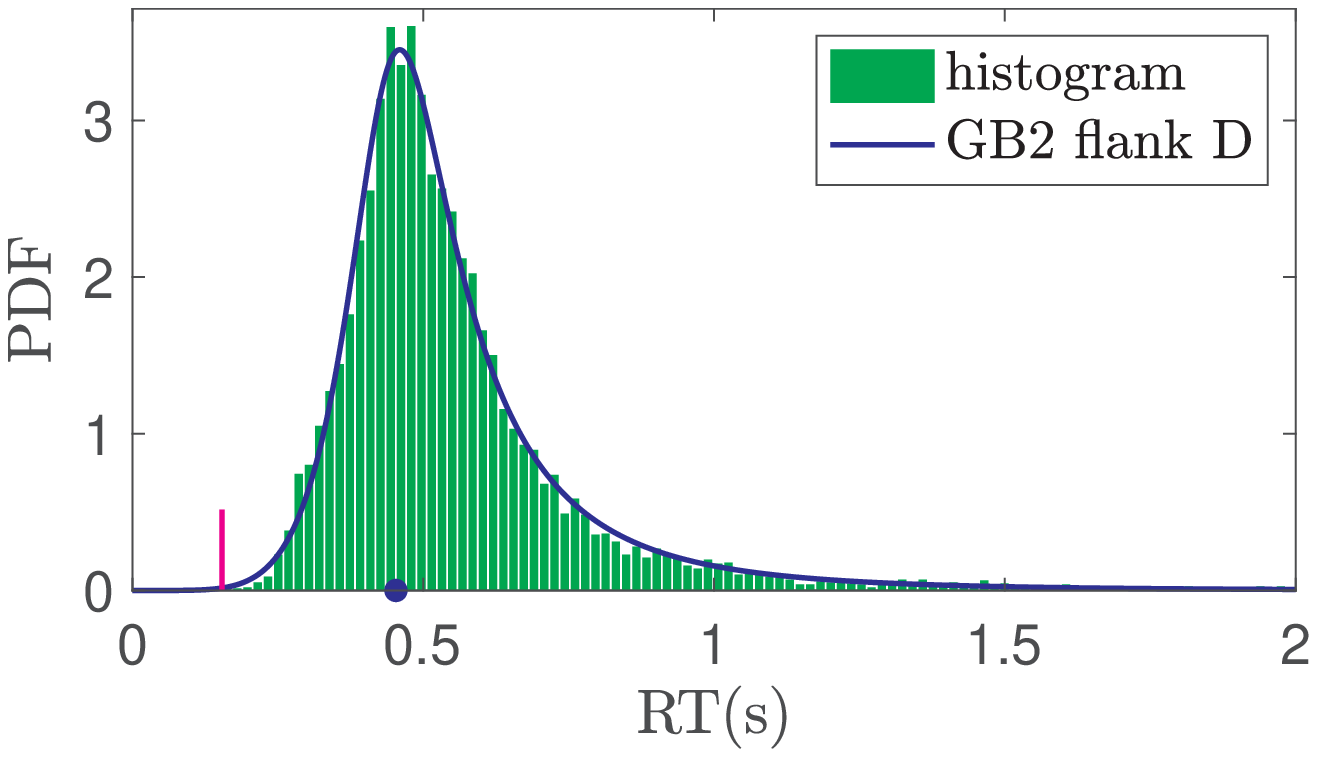}  \\ \\ \\ \\ \\
\includegraphics[width = \myFigureWidth \textwidth]{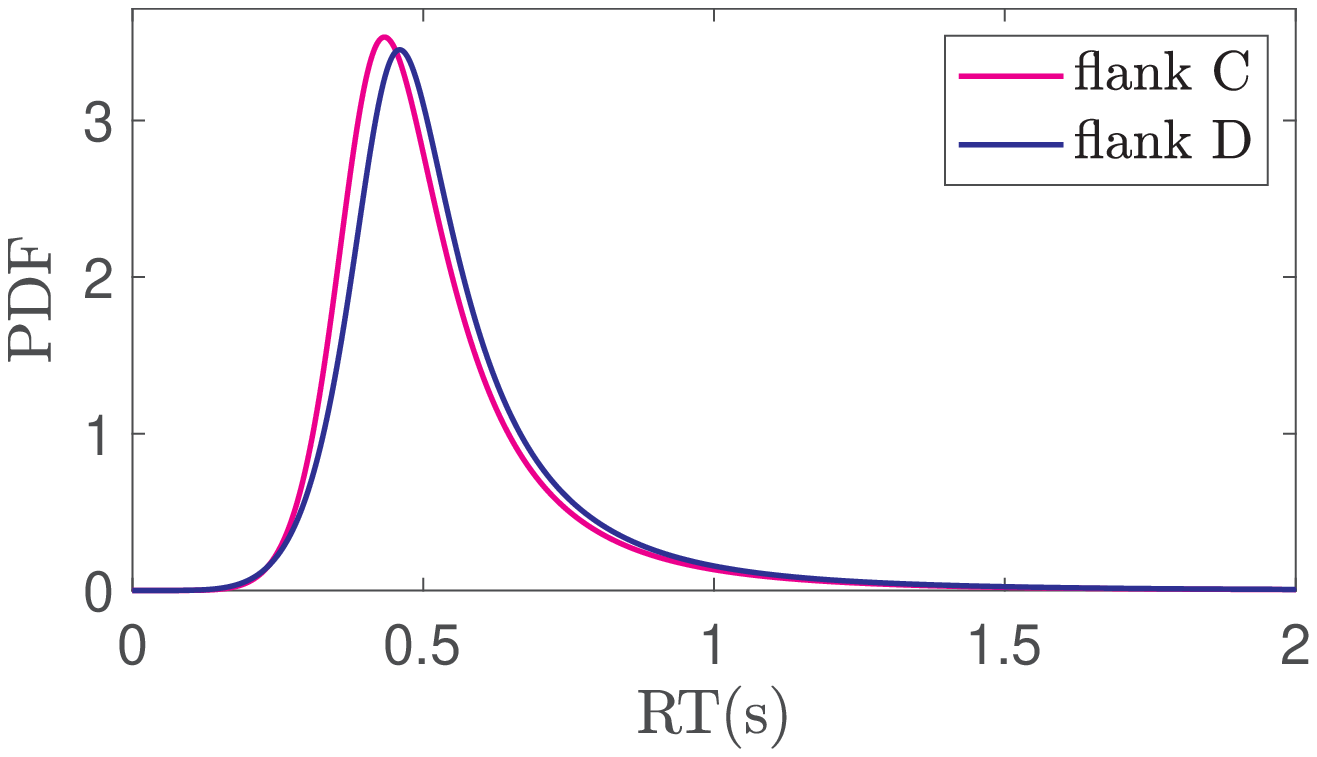}  \\ \\ \\ \\
\includegraphics[width = \myFigureWidth \textwidth]{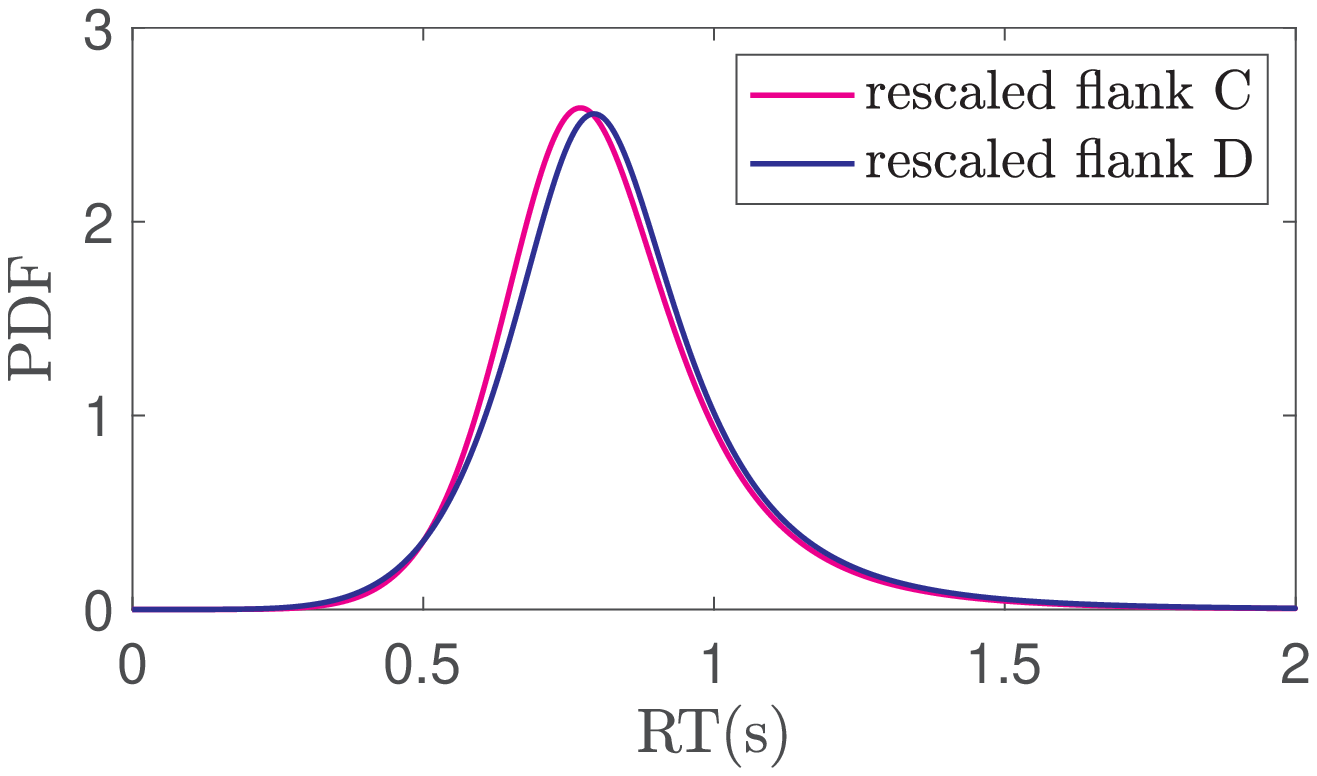} \\
\end{tabular}
\caption{Flanker test GB2 fits. From top to bottom: fit of control
group, fit of dyslexic group, fits of two top
compared, the fit above rescaled to unity mean. Vertical bar (red) shows data cut-off, black dot shows the mean.}
\label{figure1}
\end{figure*}

\begin{figure*}[!htbp]
\centering
\begin{tabular}{cc}
\includegraphics[width = \myFigureWidth \textwidth]{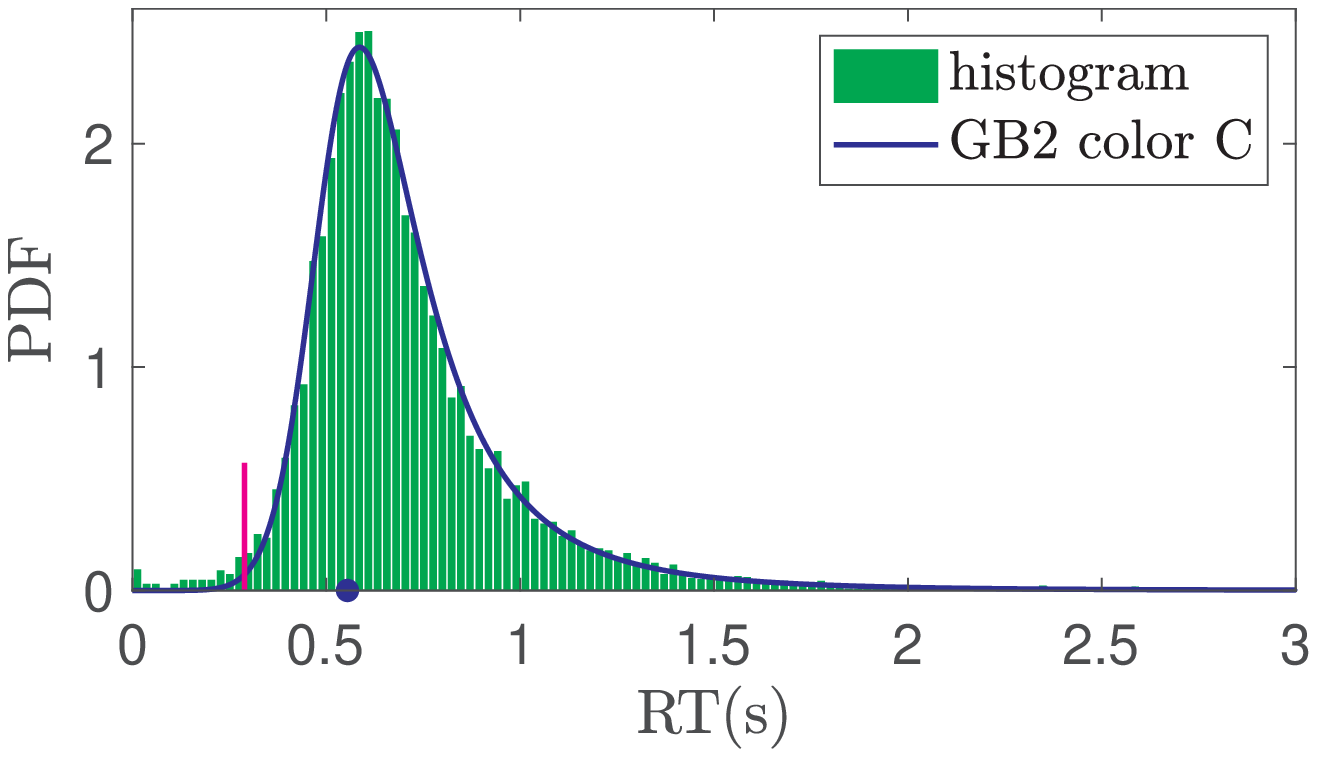} \\ \\ \\ \\ \\
\includegraphics[width = \myFigureWidth \textwidth]{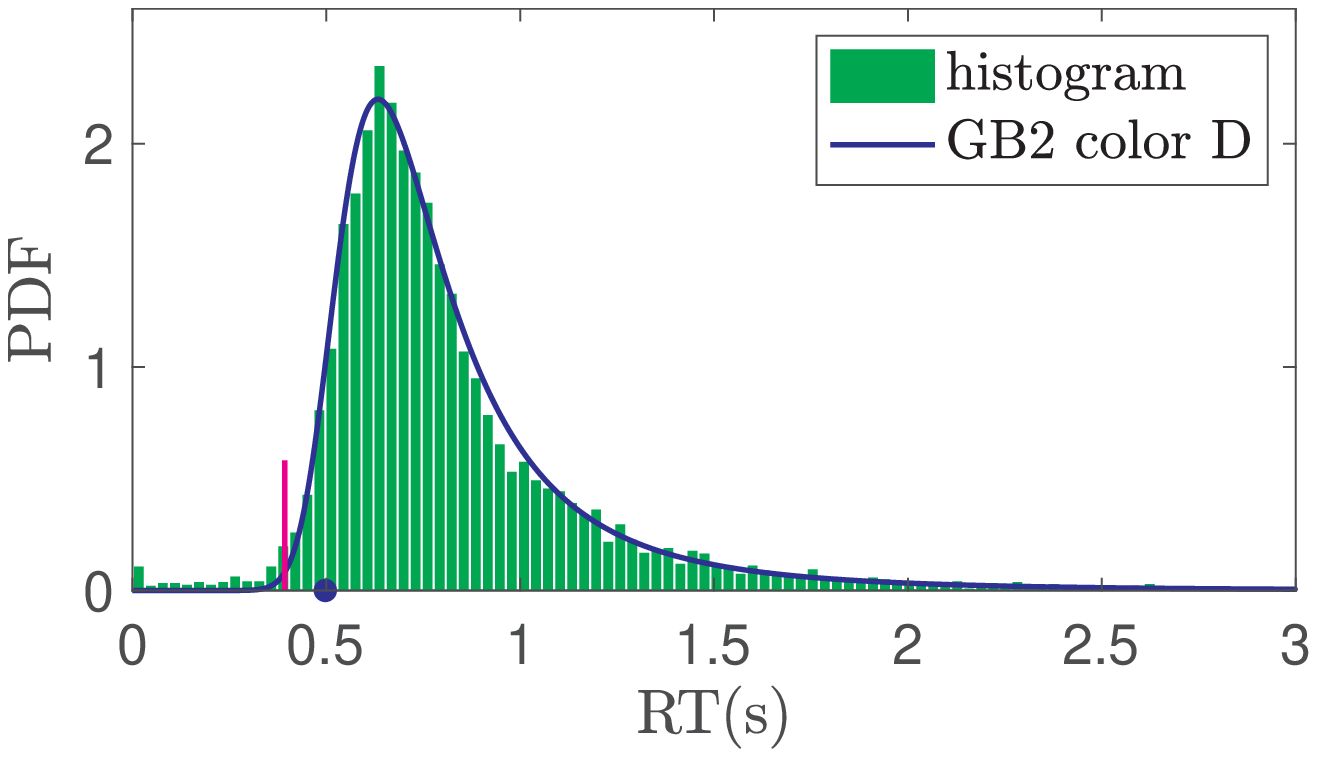} \\ \\ \\ \\ \\
\includegraphics[width = \myFigureWidth \textwidth]{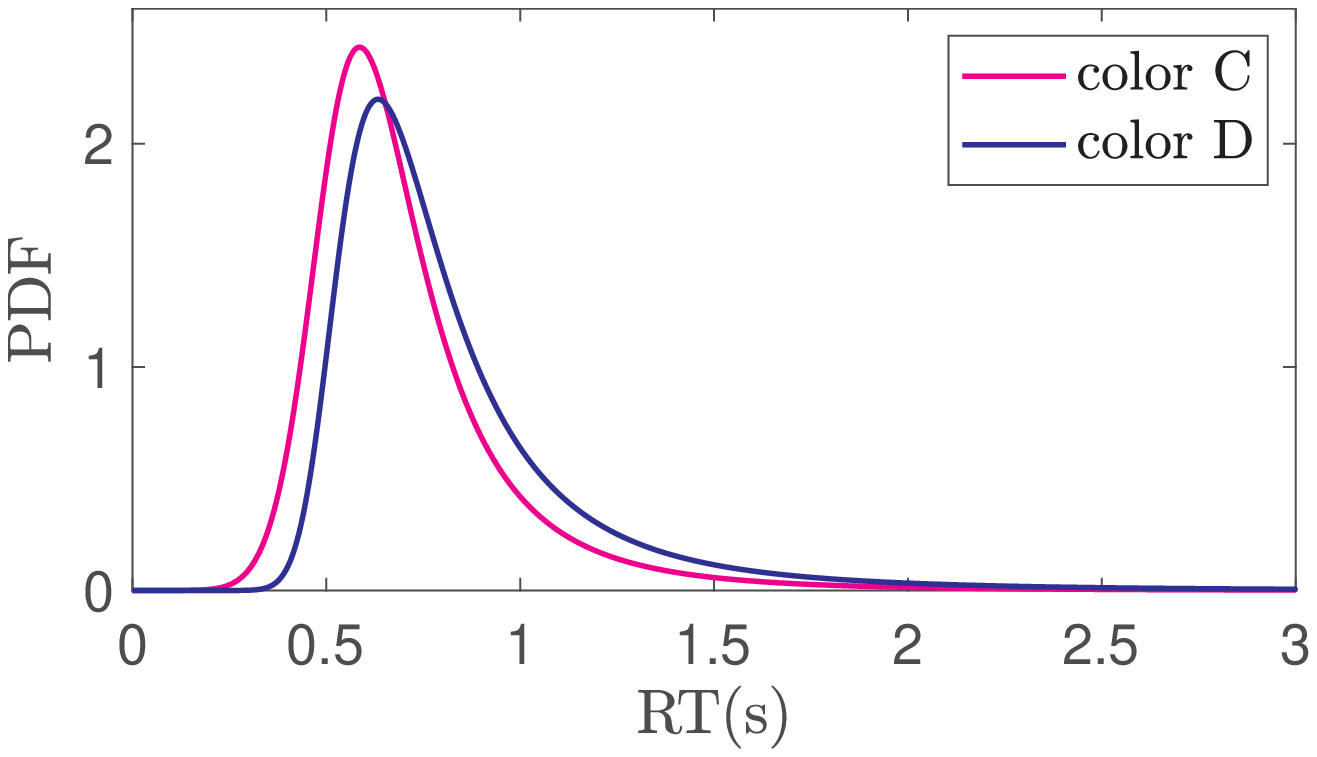} \\ \\ \\ \\
\includegraphics[width = \myFigureWidth \textwidth]{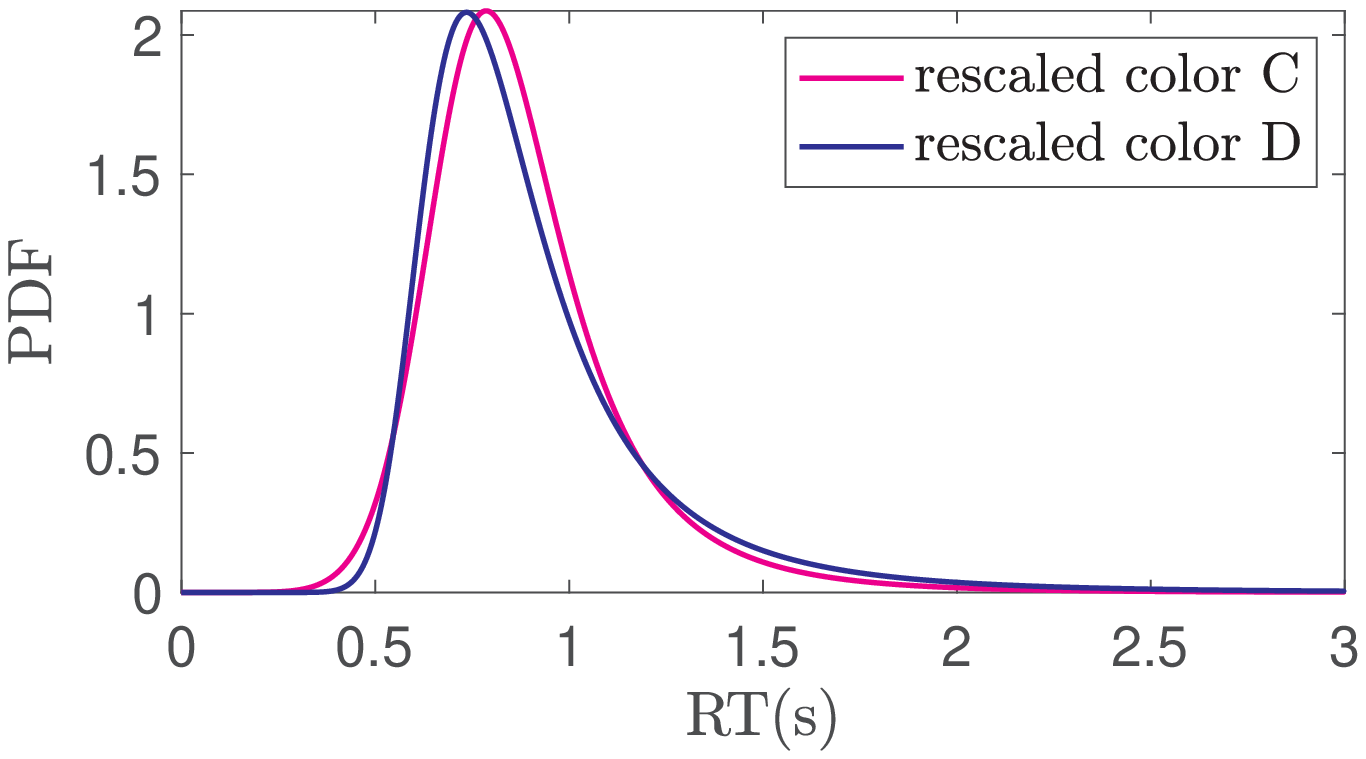}\\ 
\end{tabular}
\caption{Color-Naming test GB2 fits. From top to bottom: fit of control
group, fit of dyslexic group, fits of two top
compared, the fit above rescaled to unity mean. Vertical bar (red) shows data cut-off, black dot shows the mean.}
\label{figure3}
\end{figure*}

\begin{figure*}[!htbp]
\centering
\begin{tabular}{cc}
\includegraphics[width = \myFigureWidth \textwidth]{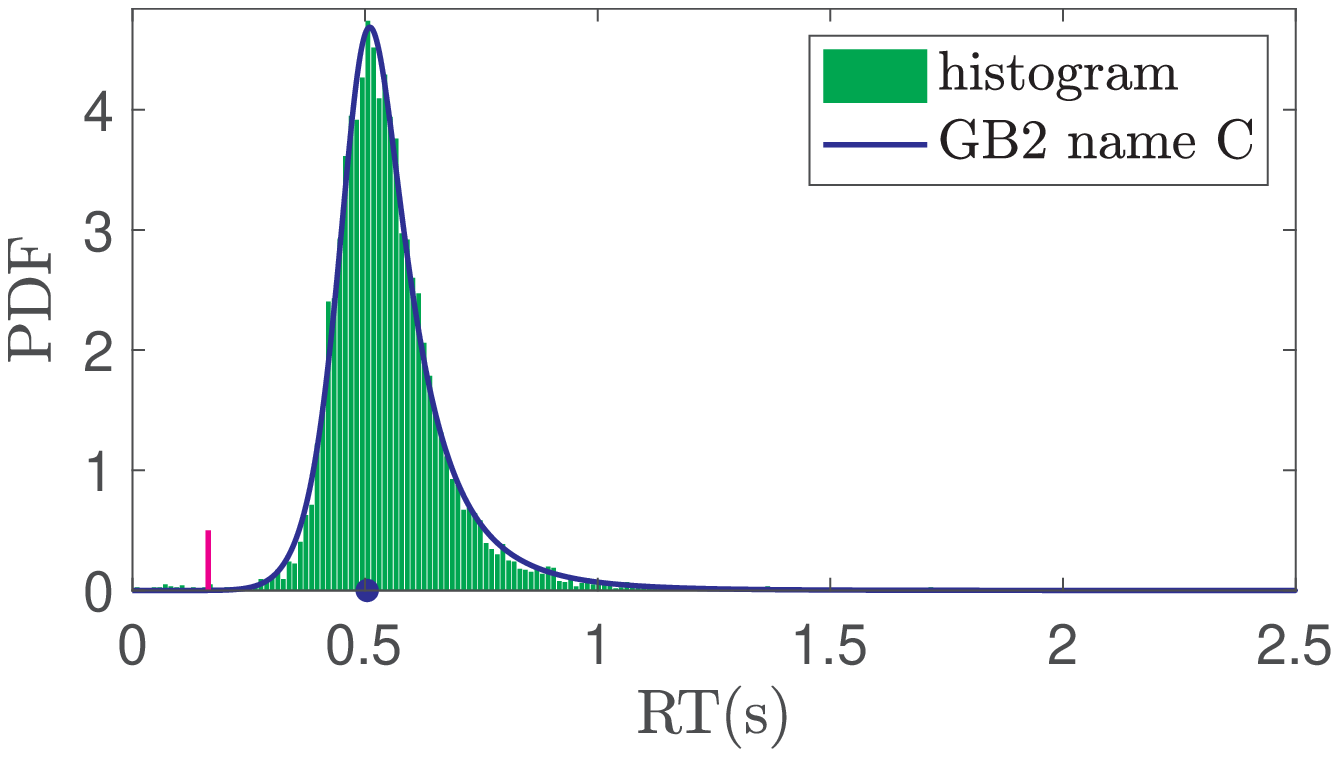}  \\ \\ \\ \\ \\
\includegraphics[width = \myFigureWidth \textwidth]{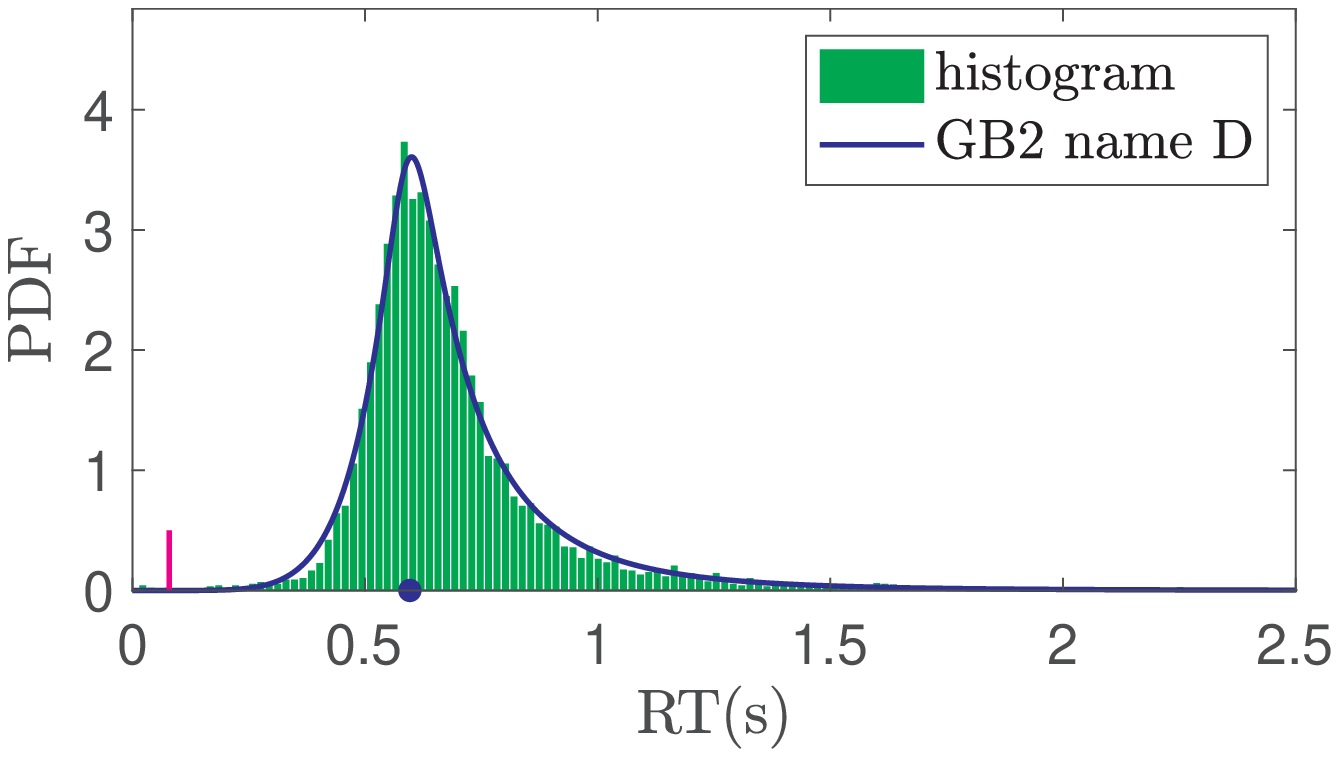}  \\ \\ \\ \\ \\
\includegraphics[width = \myFigureWidth \textwidth]{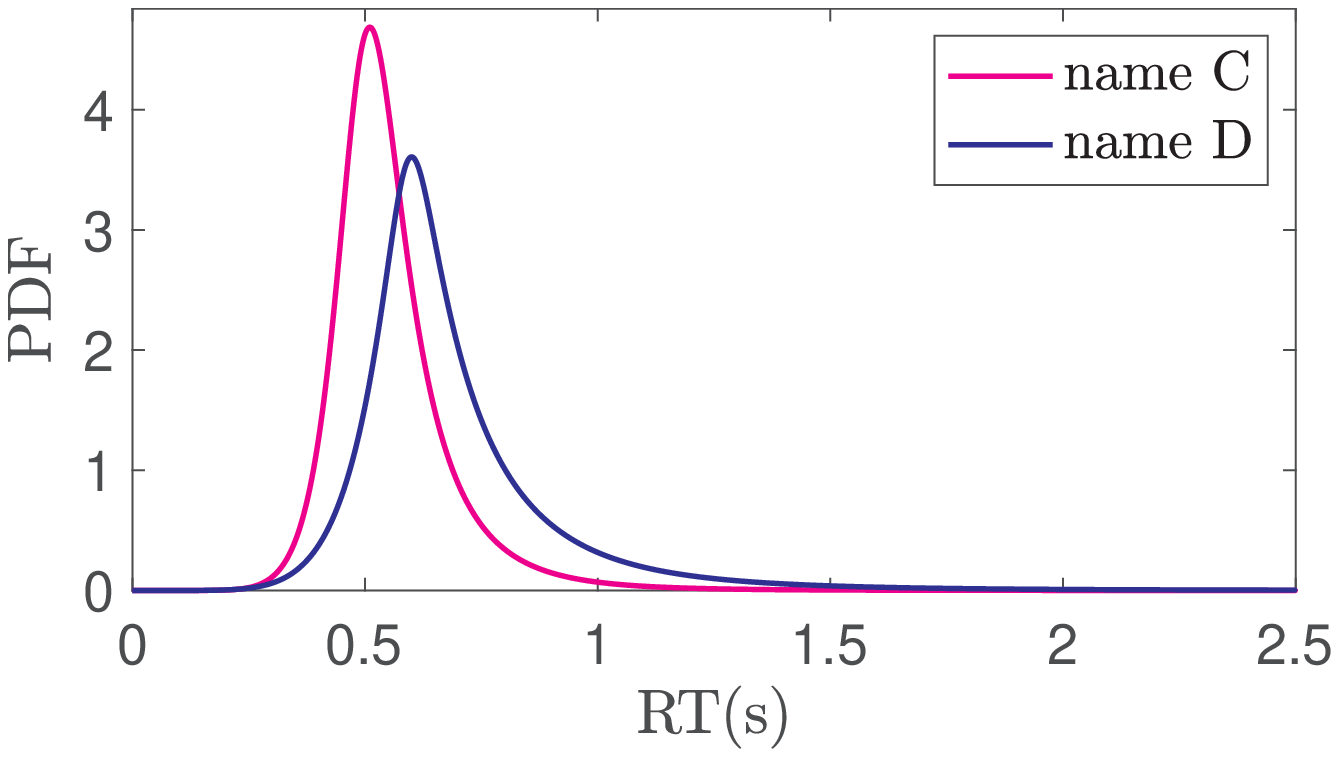}  \\ \\ \\ \\
\includegraphics[width = \myFigureWidth \textwidth]{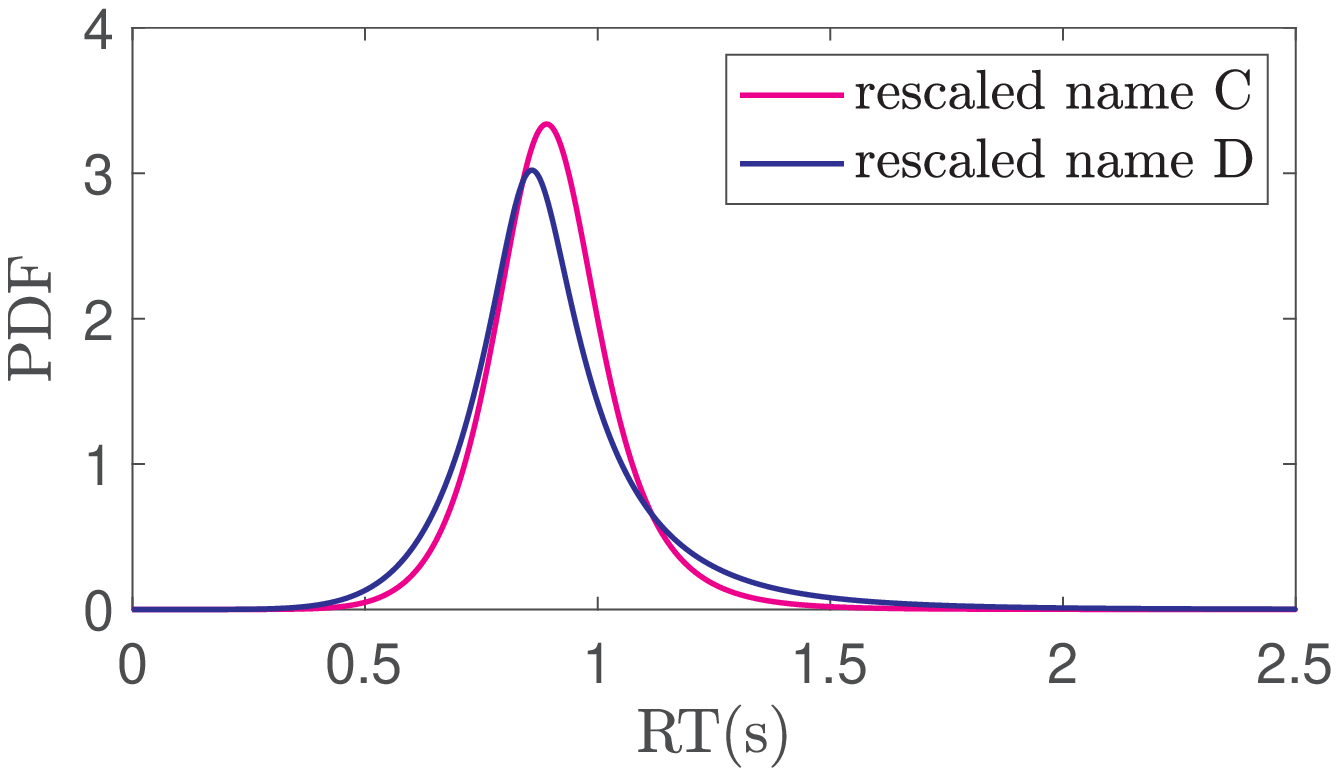}  \\
\end{tabular}
\caption{Word-Naming test GB2 fits. From top to bottom: fit of control
group, fit of dyslexic group, fits of two top
compared, the fit above rescaled to unity mean. Vertical bar (red) shows data cut-off, black dot shows the mean.}
\label{figure2}
\end{figure*}

\begin{figure*}[!htbp]
\centering
\begin{tabular}{cc}
\includegraphics[width = \myFigureWidth \textwidth]{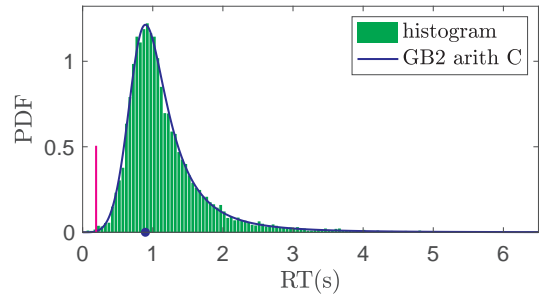} \\ \\ \\ \\ \\
\includegraphics[width = \myFigureWidth \textwidth]{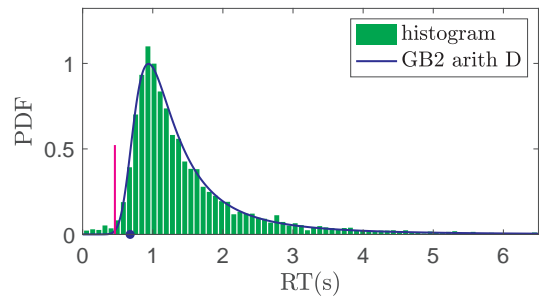}  \\ \\ \\ \\ \\
\includegraphics[width = \myFigureWidth \textwidth]{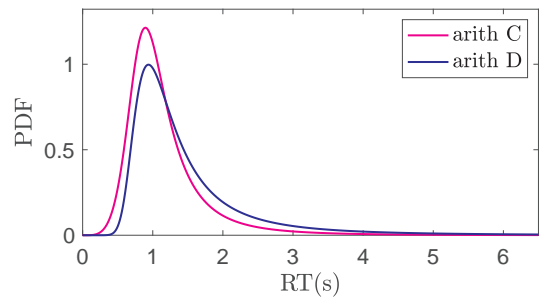} \\ \\ \\ \\
\includegraphics[width = \myFigureWidth \textwidth]{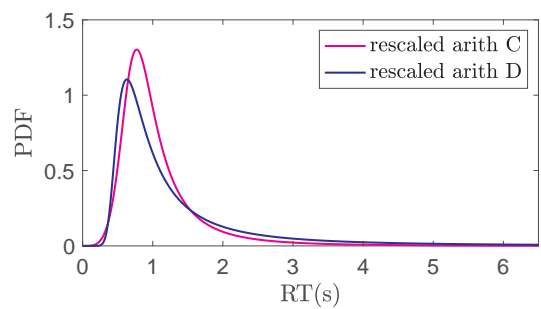}\\
\end{tabular}
\caption{Arithmetic test GB2 fits. From top to bottom: fit of control
group, fit of dyslexic group, fits of two top
compared, the fit above rescaled to unity mean. Vertical bar (red) shows data cut-off, black dot shows the mean.}
\label{figure4}
\end{figure*}

Table \ref{KSArithControl} shows results of MLE fitting for distributions mentioned in the Introduction. Lower KS numbers correspond to better fits. Clearly, fat-tailed distributions perform considerably better and, of those, GB2 is by far the best. In the rest of the paper, therefore, we concentrate solely on GB2. Figs. \ref{figure1}-\ref{figure4} show RT distributions fits with GB2 and their comparison between control and dyslexic groups, including rescaled versions in order to emphasize shape differences or lack thereof. We will use bootstrap in Section \ref{BootstrapCI} to more precisely identify differences between shape-related quantities introduced in Section \ref{New}. Table \ref{GB2cut} gives parameters of GB2 fits and the values of front and tail exponents.
\begin{table}[!htbp]
\centering
\caption{KS statistics for distributions in Introduction. Lower numbers correspond to better fits.}
\label{KSArithControl}
\begin{tabular}{ccccccc} 
\hline
\hline
            &    GB2 &    LNP &    GIGa &    GIGa3 &    exGaussian &    weibull3\\
\hline
    FlankControl &    0.008   &     0.008   &	0.020	& 	0.025	&	0.039	&     0.072 \\
\hline
    FlankDyslexia &   0.012  &      0.015  &	0.034	& 	0.039	&	0.055	&     0.091 \\
\hline
    ColorControl &  0.008 & 0.019 &  0.029	& 0.032 &	0.032 & 0.072    \\
\hline
    ColorDyslexia & 0.010 & 	0.017 & 0.025 & 0.030 &	0.047 & 0.076   \\
\hline
   NameControl &  0.010  &    0.012   & 0.026 & 0.029	&	0.029	&  0.066  \\
\hline
    NameDyslexia & 0.014 &   0.014  & 0.035 & 	0.040	&	0.057	&  0.090    \\
    \hline
  ArithControl &      0.007 & 0.015 & 0.022 & 0.025 & 0.032 & 0.067 \\
\hline
    ArithDyslexia &  0.013 & 0.019 & 0.026 & 0.028	 &	0.068 & 0.056  \\
\hline
\hline
\end{tabular}
\end{table}

\begin{table}[!htbp]
\centering
\caption{GB2 fitting parameters and front $\alpha p - 1$ and tail $-(\alpha q + 1)$ exponents.}
\label{GB2cut}
\begin{tabular}{ccccc}
\hline
\hline
		&	parameters&	   front exponent &    tail exponent & KS test \\
\hline
FlankControl& (          0.7986,           0.3768,		9.6961,           0.4169) &           6.7434 &          -4.6535 &           0.008 \\
\hline
FlankDyslexia& (          0.4992,           0.2755,		12.9841,           0.4530) &           5.4818 &          -4.5767 &           0.012 \\
\hline
ColorControl& (          1.2003,           0.5783,           6.8557,          0.5545) &           7.2288 &          -4.9646 & 0.008  \\
\hline
ColorDyslexia& (          3.2487,           0.5668,           5.9343,          0.4975) &          18.2787 &          -4.3635 & 0.010  \\
\hline
NameControl& (          0.6104,           0.3957,           15.6084,          0.5045) &           8.5273 &          -7.1762  & 0.010\\
\hline
NameDyslexia& (          0.2398,           0.1380,           30.5193,          0.5963) &           6.3185 &          -5.2116 & 0.014  \\
\hline
AirthControl& (          0.8211,           0.4962,           6.1620,           0.8949) &           4.0596 &          -4.0575  & 0.007\\
\hline
AirthDyslexia&(          3.2780,           0.5248,           4.2612,          0.6777) &           12.9682 &          -3.2362 & 0.013 \\
\hline\hline
\end{tabular}
\end{table}

\clearpage

\subsection{Measures of Variability \label{New}} 
\begin{figure}[!htbp]
\centering
\begin{tabular}{cc}
\includegraphics[width = 0.47 \textwidth]{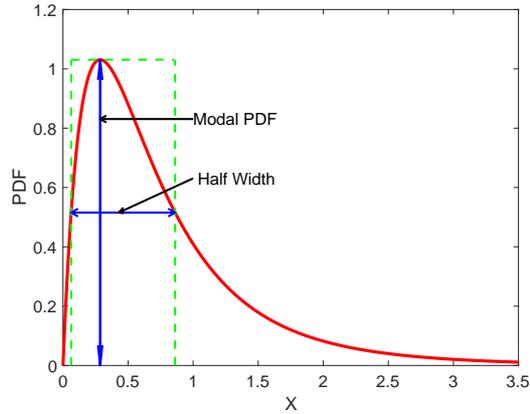}
\end{tabular}
\caption{Illustration of DMMS. Area inside represents roughly the proportion of subjects not in the front end and tails. DMMS represents roughly the proportion of subjects in the front end and tails.}
\label{Gini2graph}
\end{figure}

Here we identify several scale-independent measures of variability which will be used to compare the distributions of control and dyslexic groups. As we mentioned early, SD overweighs influence of fat tails. Therefore we utilize a different measure - Half Width. It is illustrated in  Fig. \ref{Gini2graph}, where ``modal PDF", $MPDF$, is the height of the PDF at the mode and "Half Width", $HW$, is the width of the distribution at half modal PDF. Since both $HW$ and the mean $\mu$ of the distribution are proportional to the scale parameter, $\propto \beta$, the reduced HW 
\begin{equation}
\label{rHW}
rHW=\frac{HW}{\mu}
\end{equation}
is scale-independent, that is depends only on shape parameters.

Since $MPDF \propto \beta^{-1}$ another scale-independent quantity, which measures the fraction of those with particularly low values and those in the tails, can be used \cite{dashti2019generalized}:
\begin{equation}
\label{Gini2}
DMMS=1- MPDF \times HW
\end{equation}

A well known scale-independent measure of variability, known as Hoover index in studies of income/wealth inequality, is given by
\begin{equation}
H =\frac{1}{2 \mu}\int_{0}^{\infty} p(x) |x - \mu |  \mathrm{d}x 
\label{Hoover}
\end{equation}
where $p(x)$ is the distributions PDF and $\mu$ is its mean.

\clearpage

\subsection{Bootstrap and Confidence Intervals\label{BootstrapCI}}
\begin{figure*}[!htbp]
\centering
\begin{tabular}{cc}
\includegraphics[width =0.25 \textwidth]{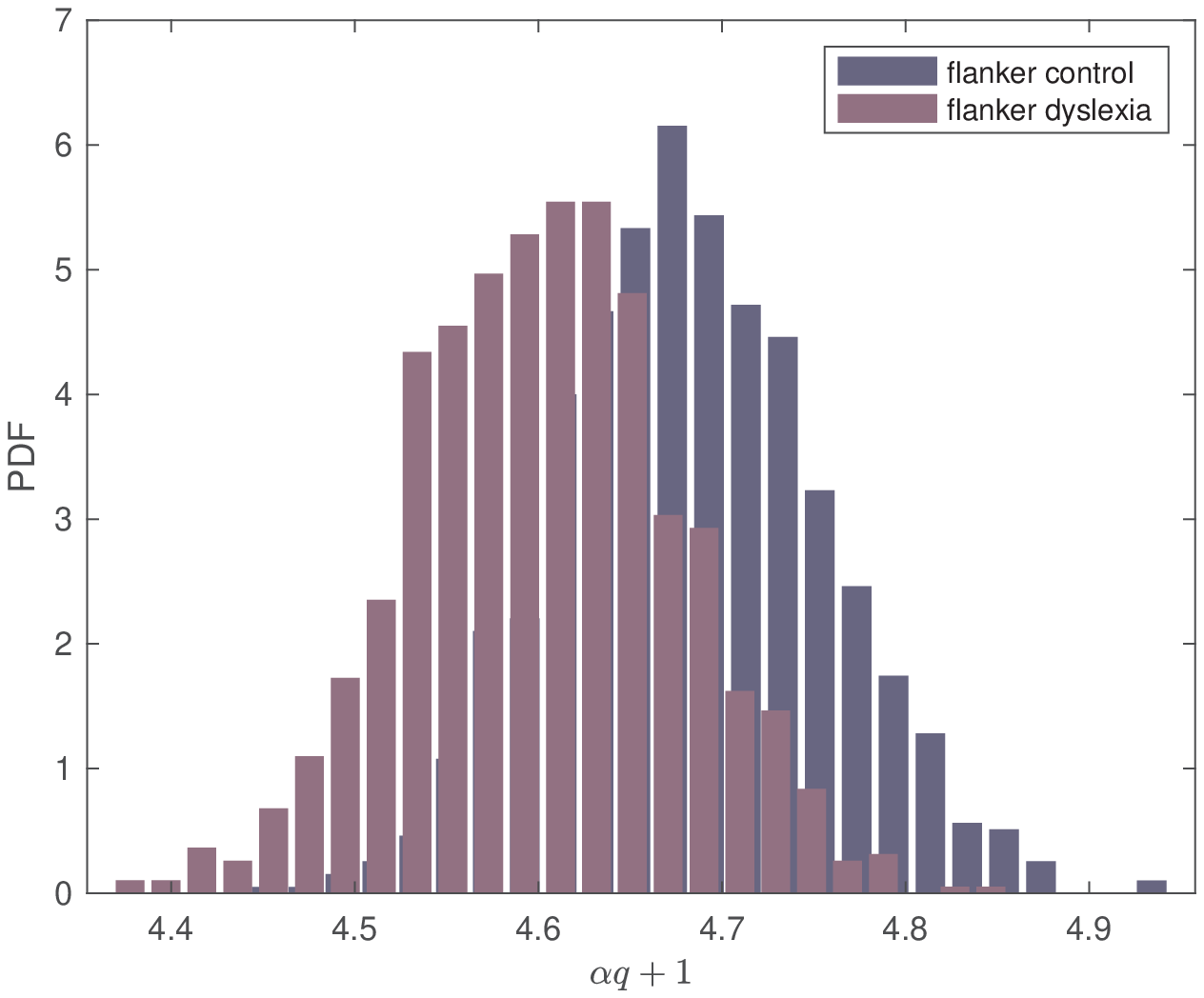}
\includegraphics[width = 0.25 \textwidth]{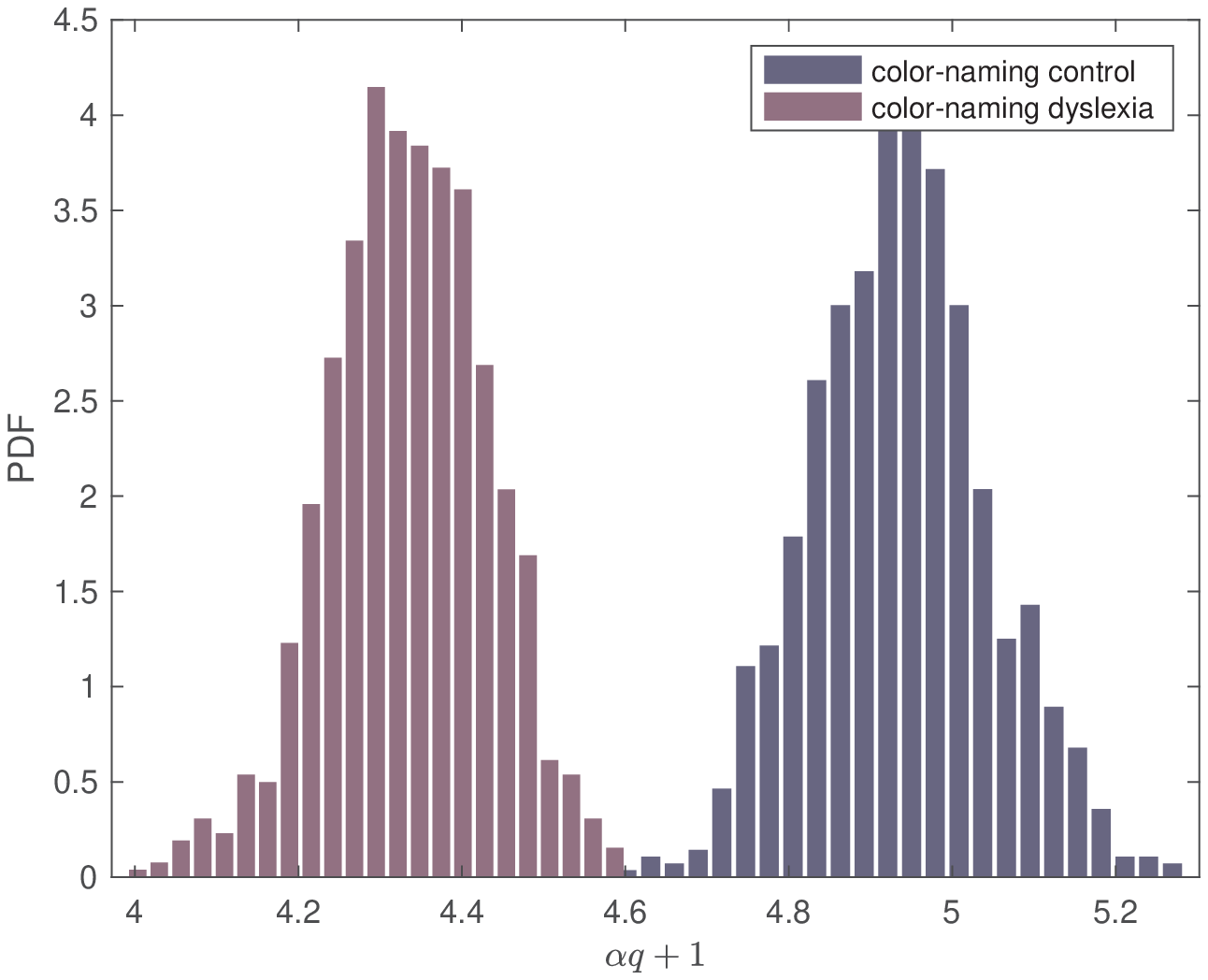} 
\includegraphics[width = 0.25 \textwidth]{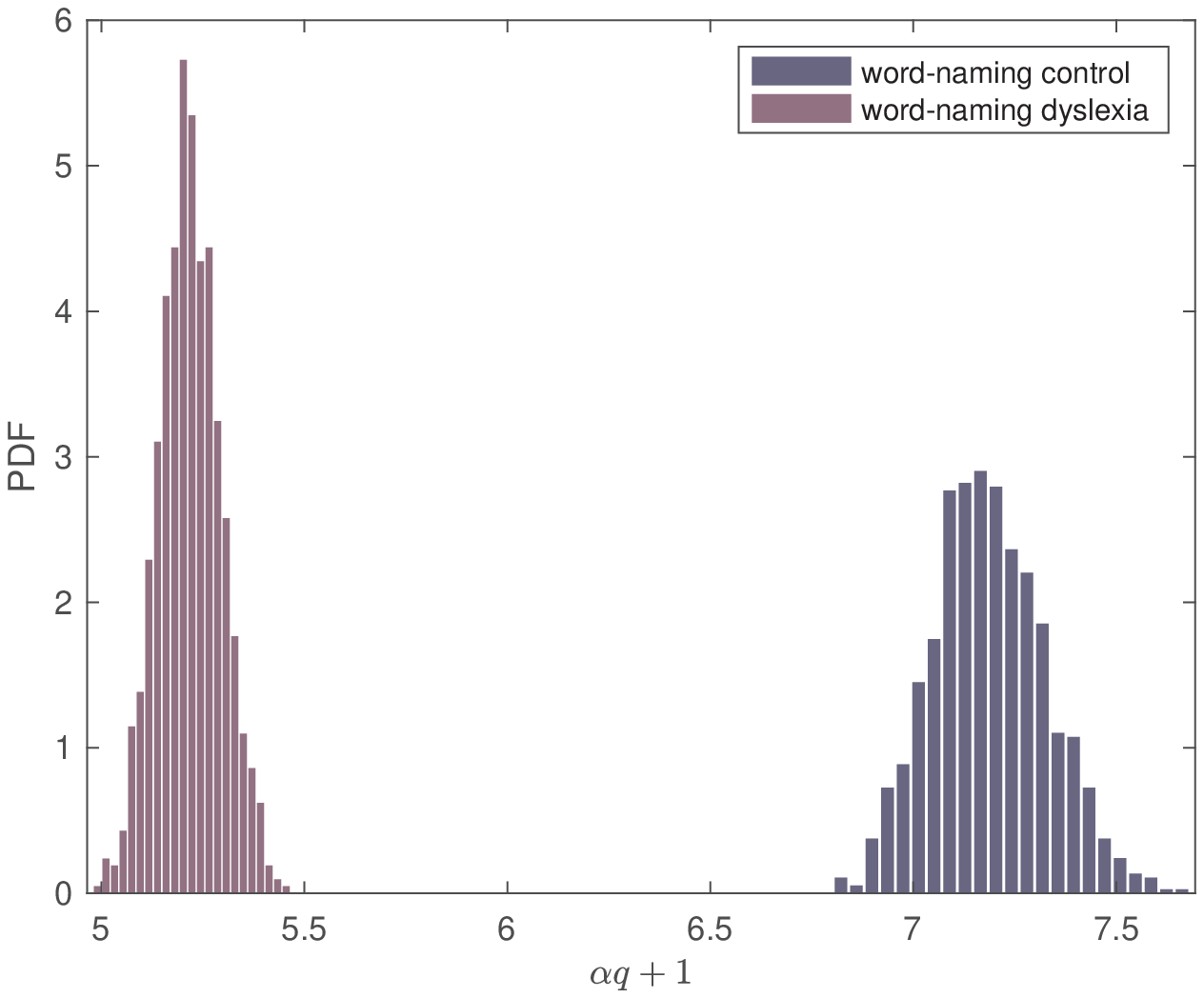}
\includegraphics[width = 0.25 \textwidth]{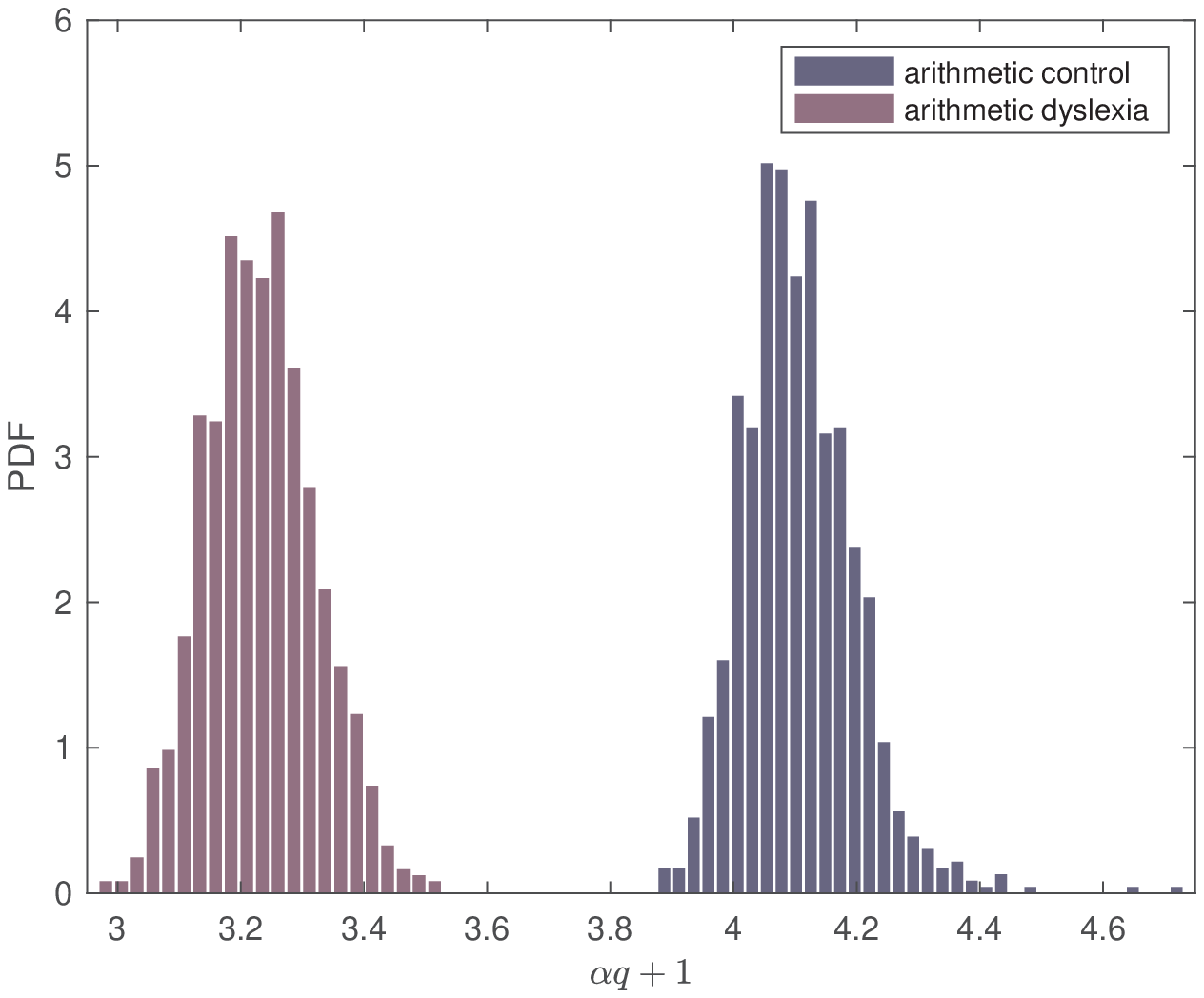} \\ \\ \\ \\
\includegraphics[width = 0.25 \textwidth]{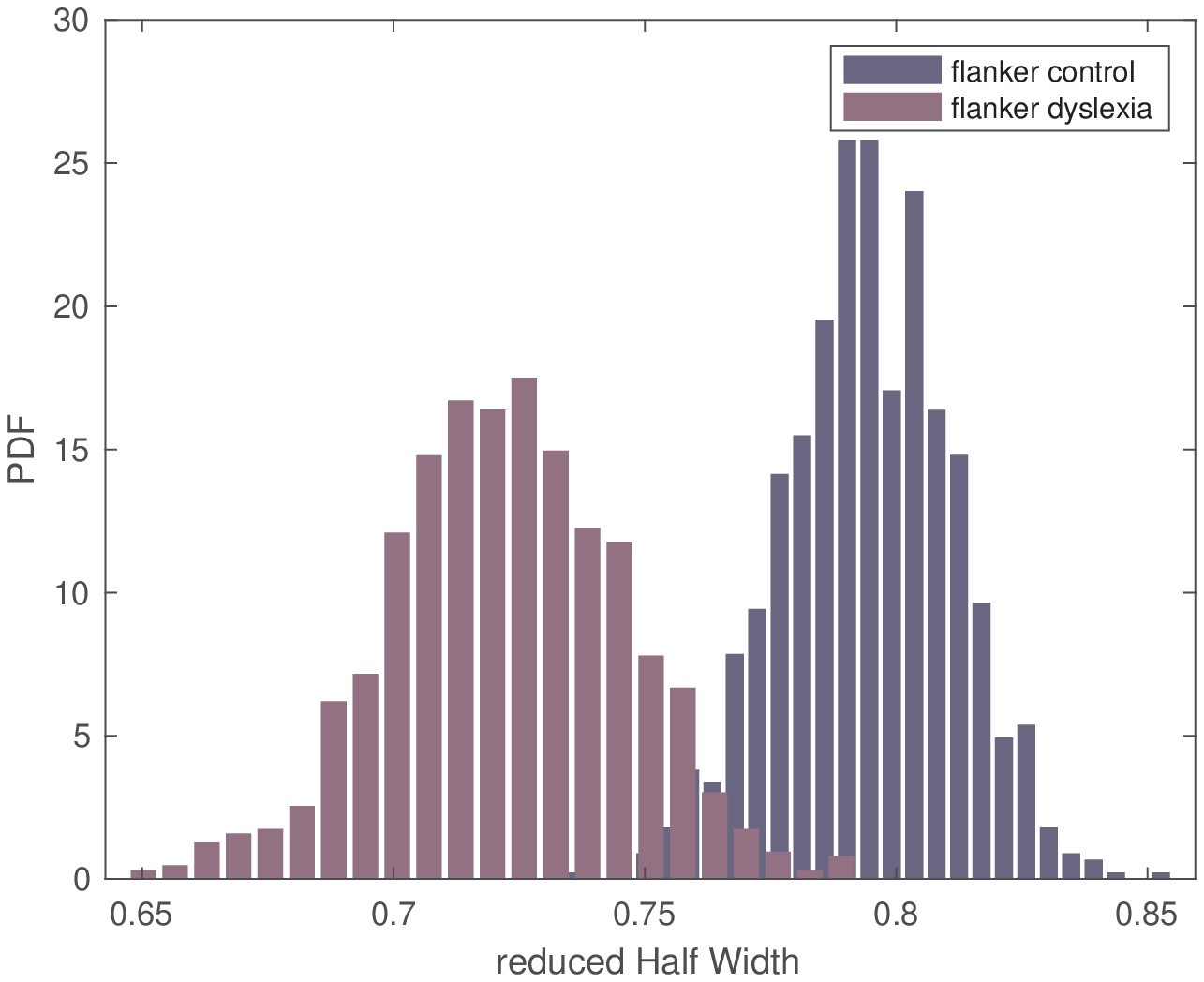}
\includegraphics[width = 0.25 \textwidth]{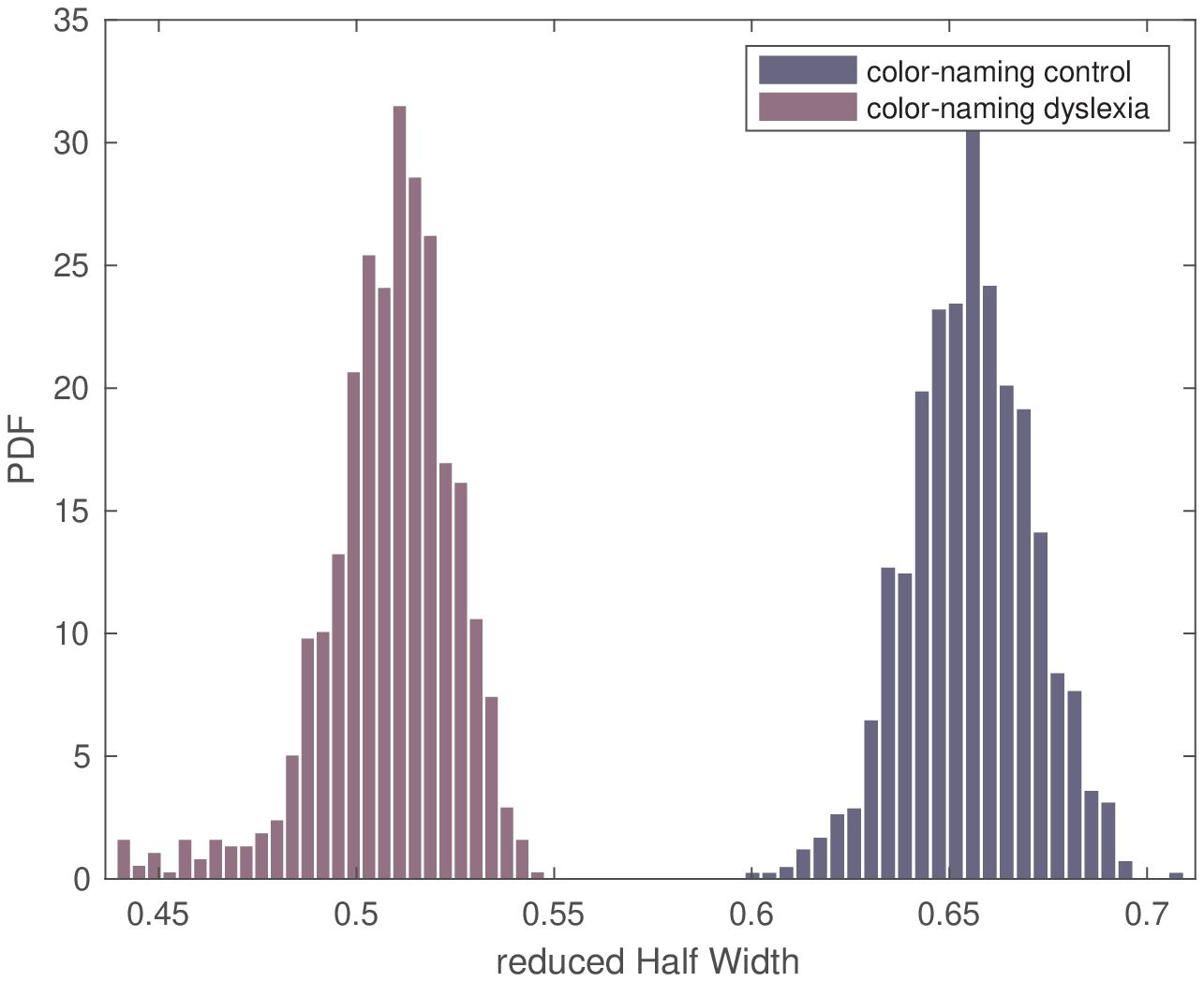}  
\includegraphics[width = 0.25 \textwidth]{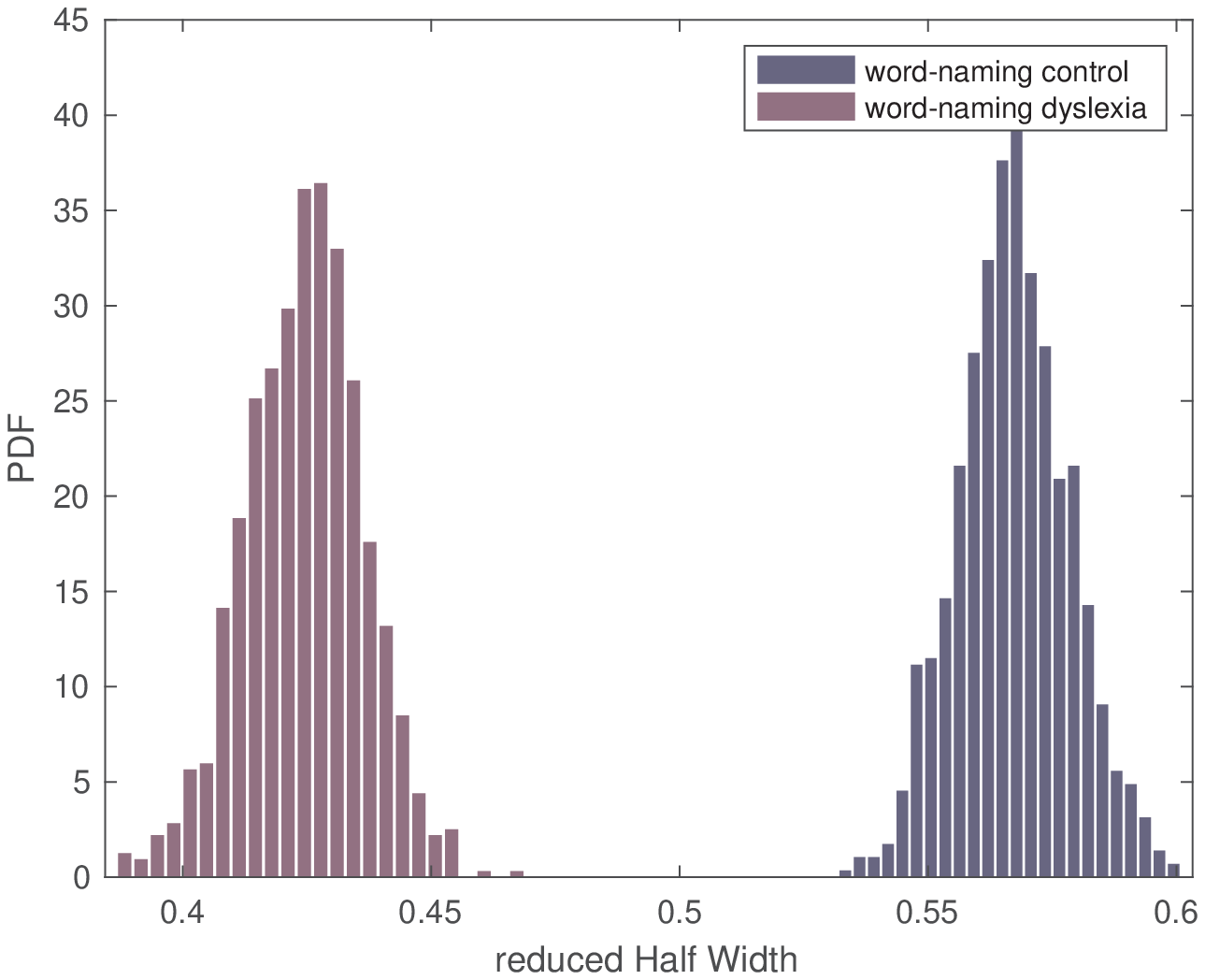}
\includegraphics[width = 0.25 \textwidth]{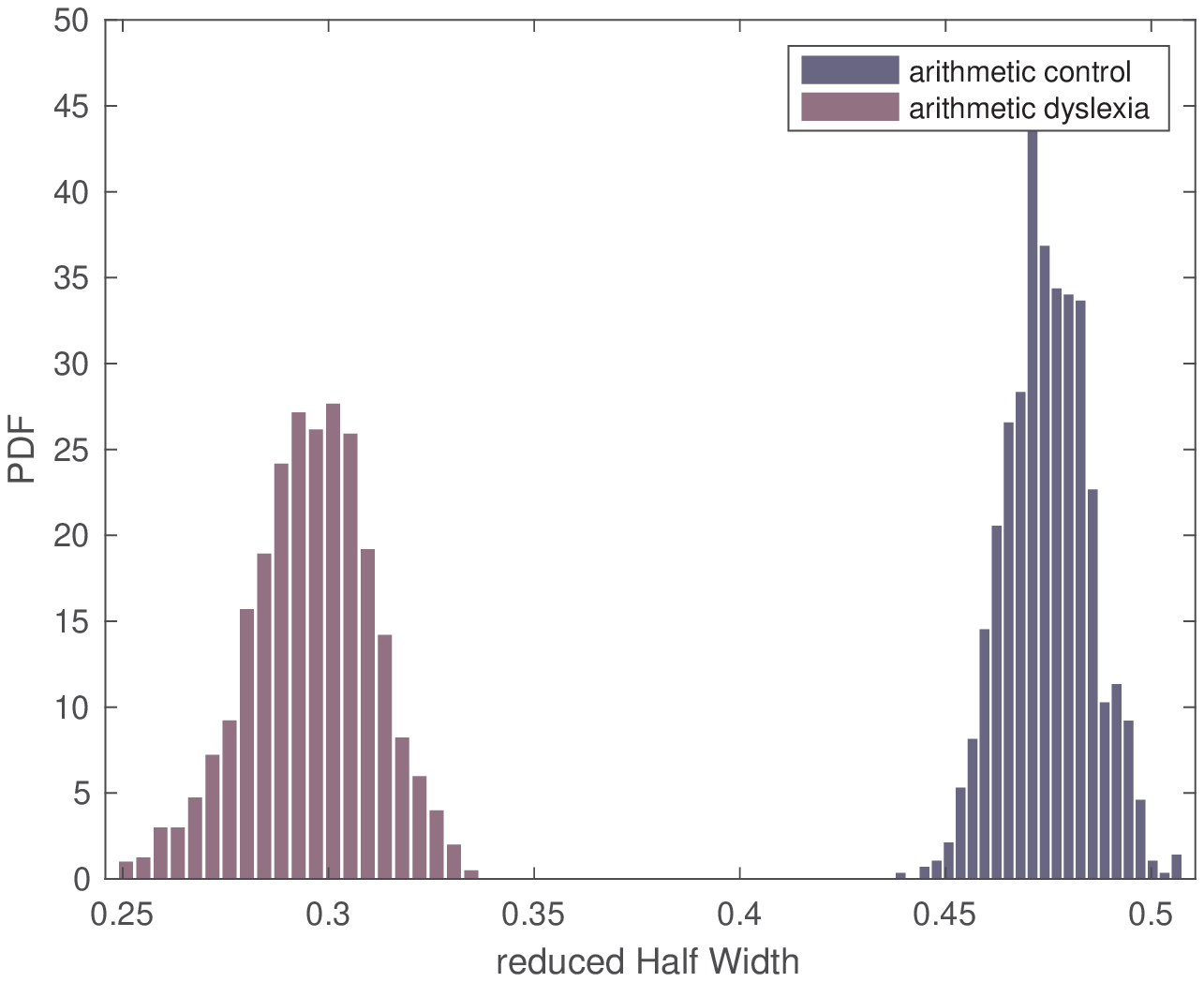} \\ \\ \\ \\
\includegraphics[width = 0.25 \textwidth]{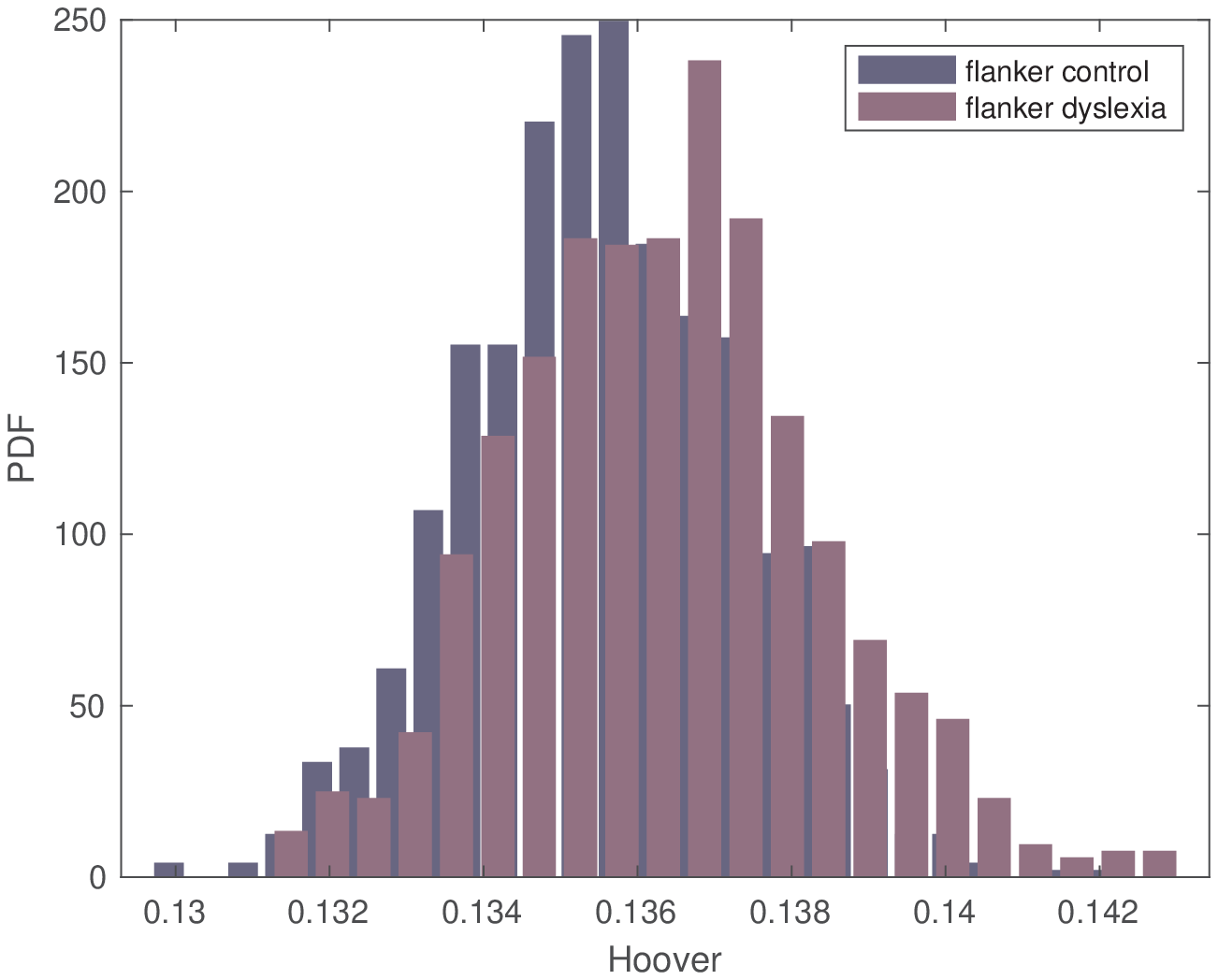} 
\includegraphics[width = 0.25 \textwidth]{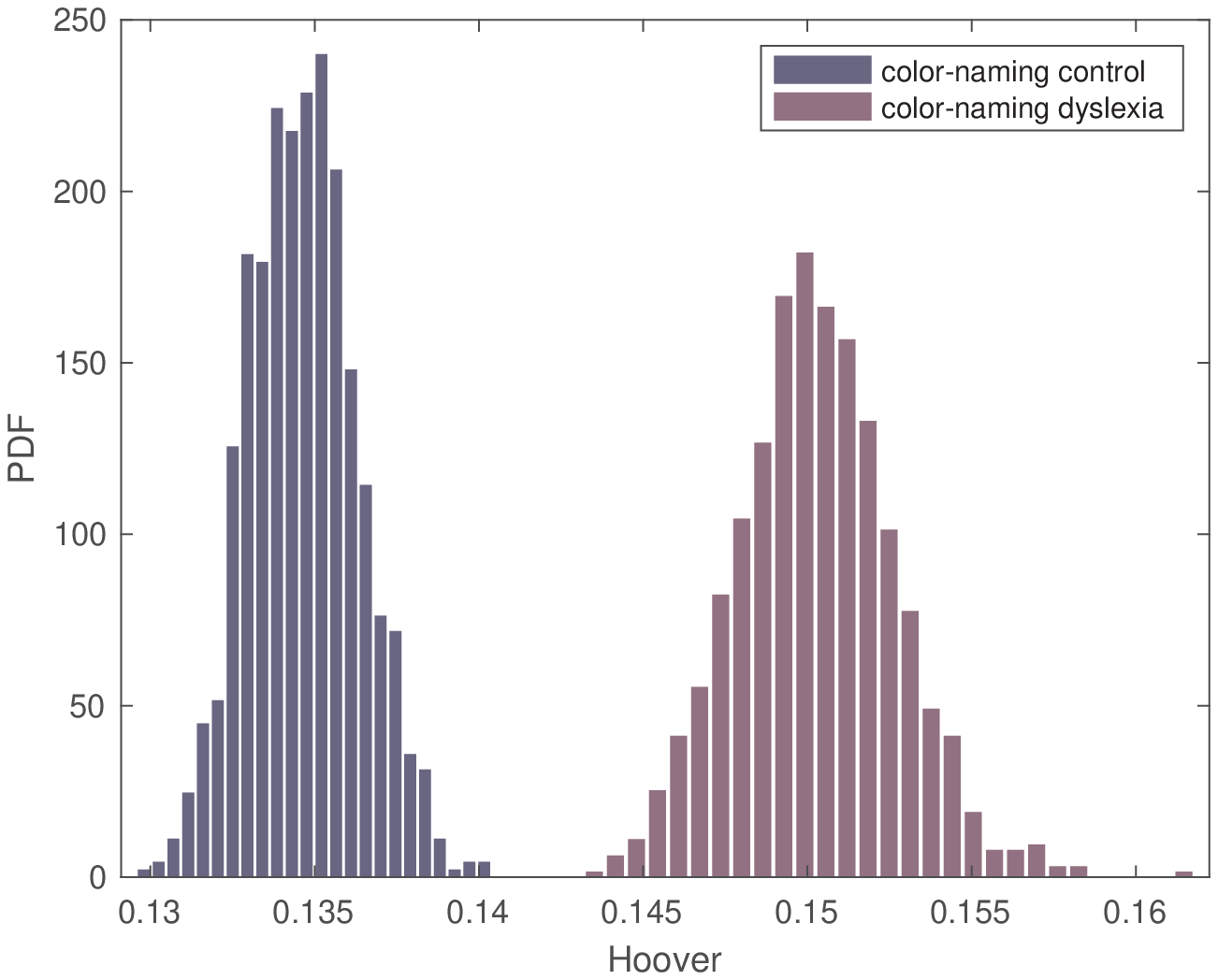}
\includegraphics[width = 0.25 \textwidth]{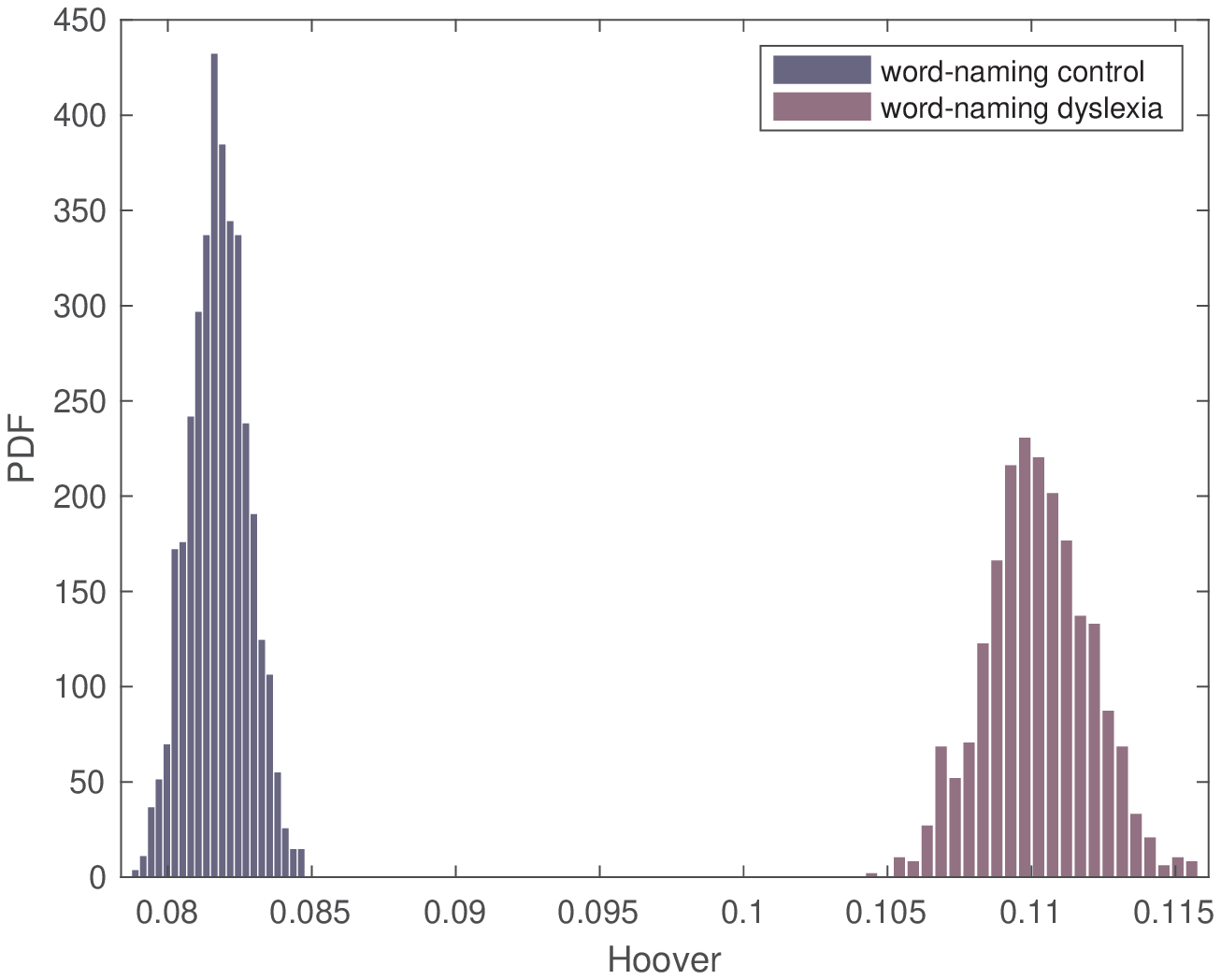}
\includegraphics[width = 0.25 \textwidth]{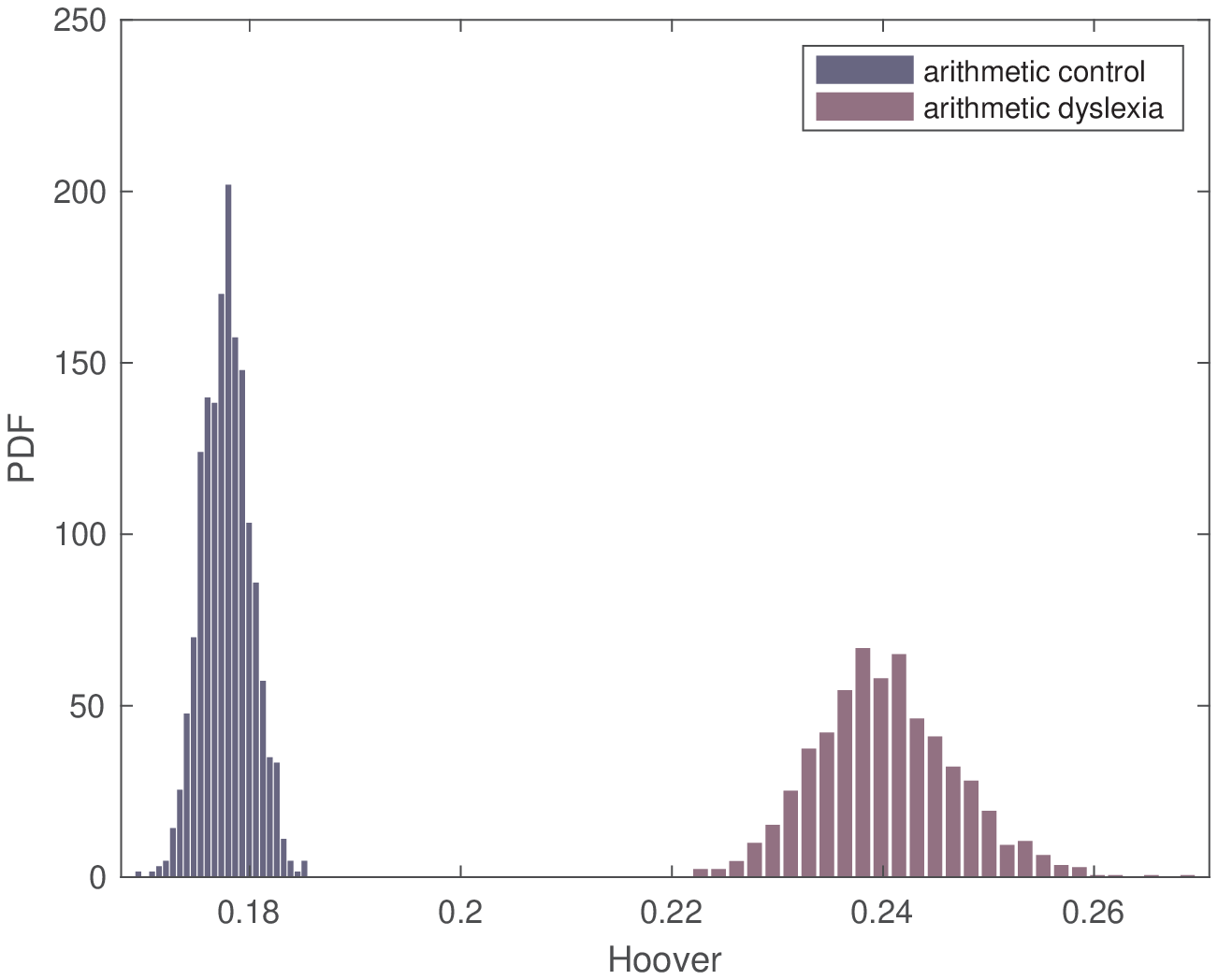} \\ \\ \\ \\
\includegraphics[width = 0.25 \textwidth]{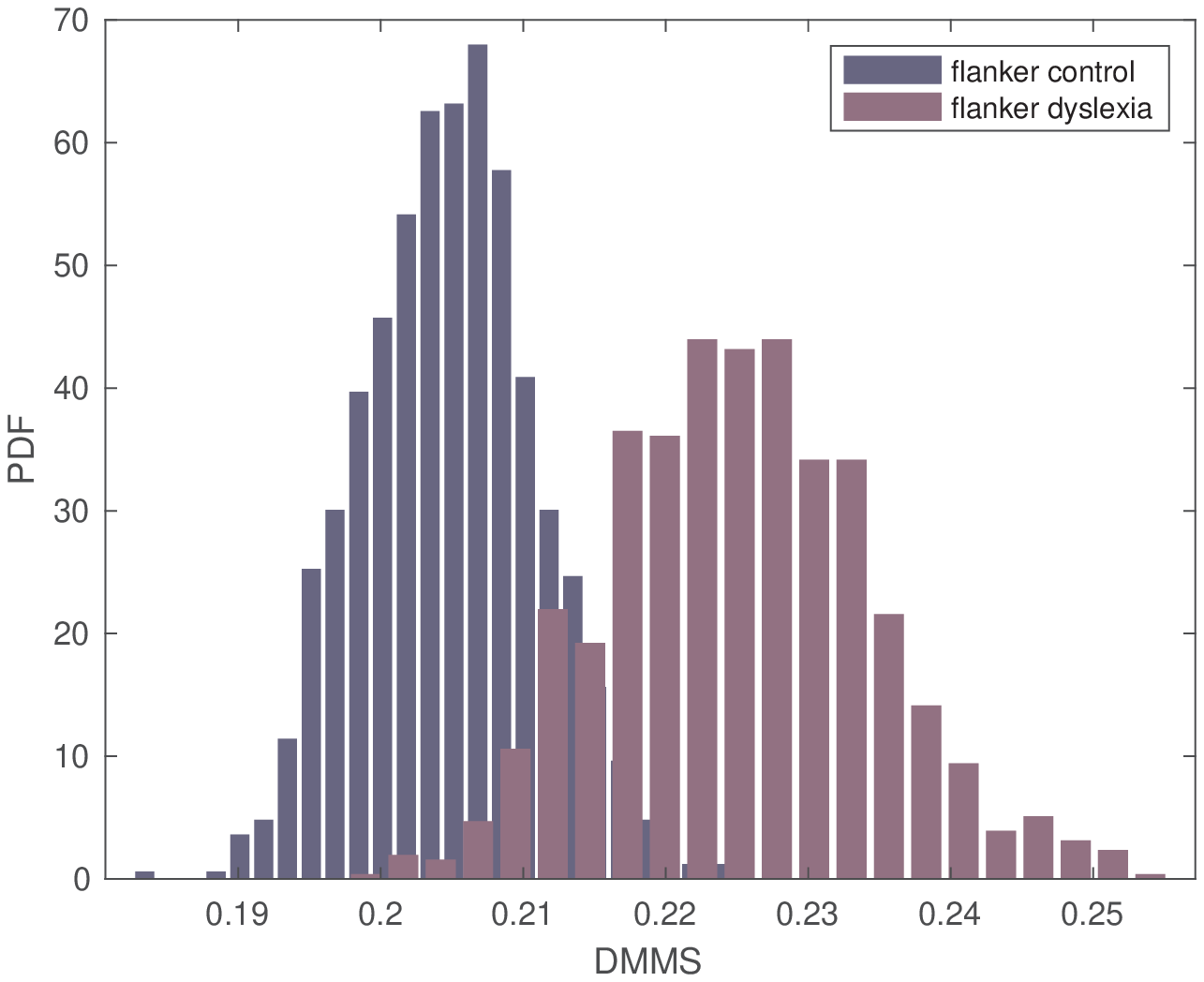} 
\includegraphics[width = 0.25 \textwidth]{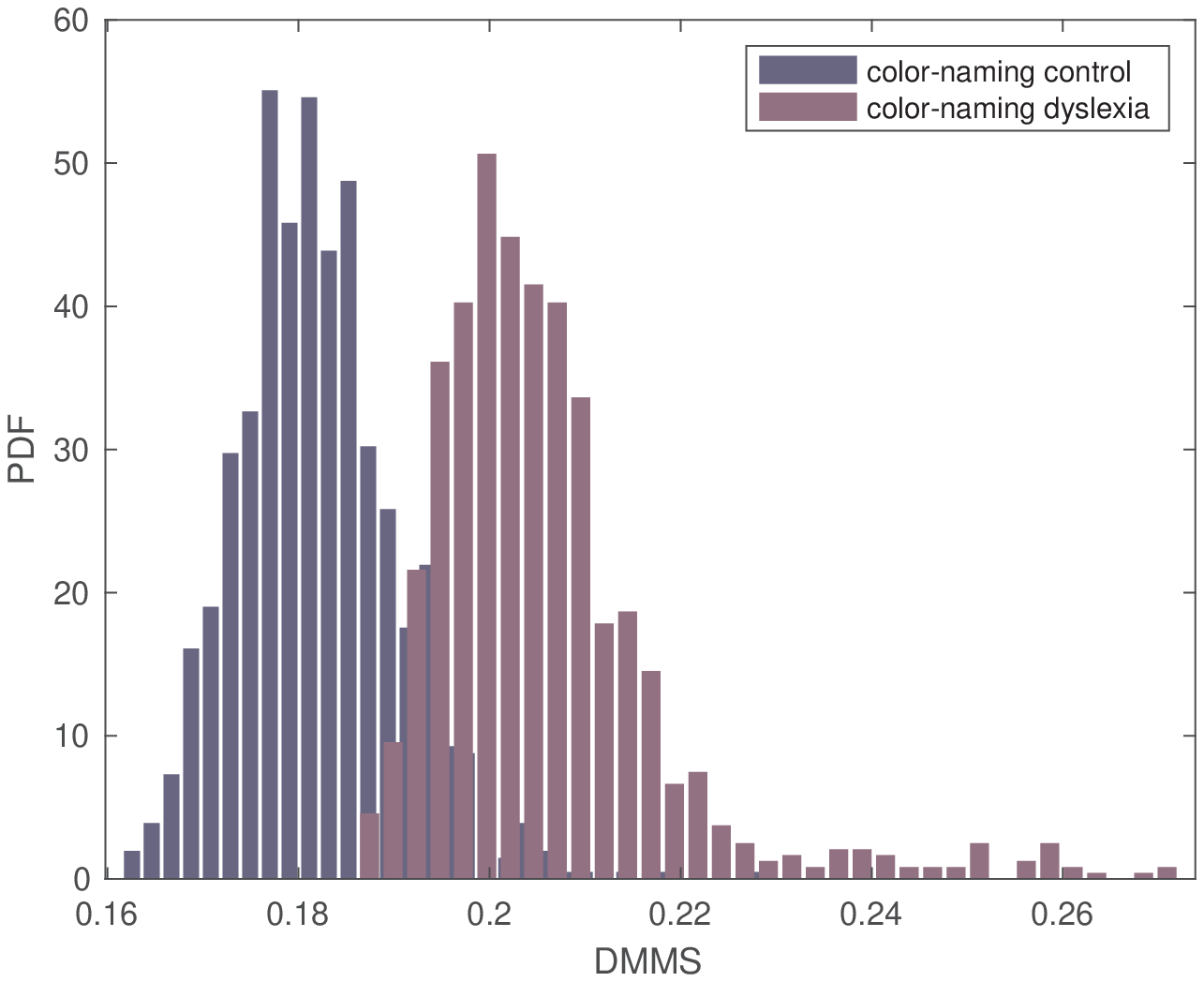}
\includegraphics[width = 0.25 \textwidth]{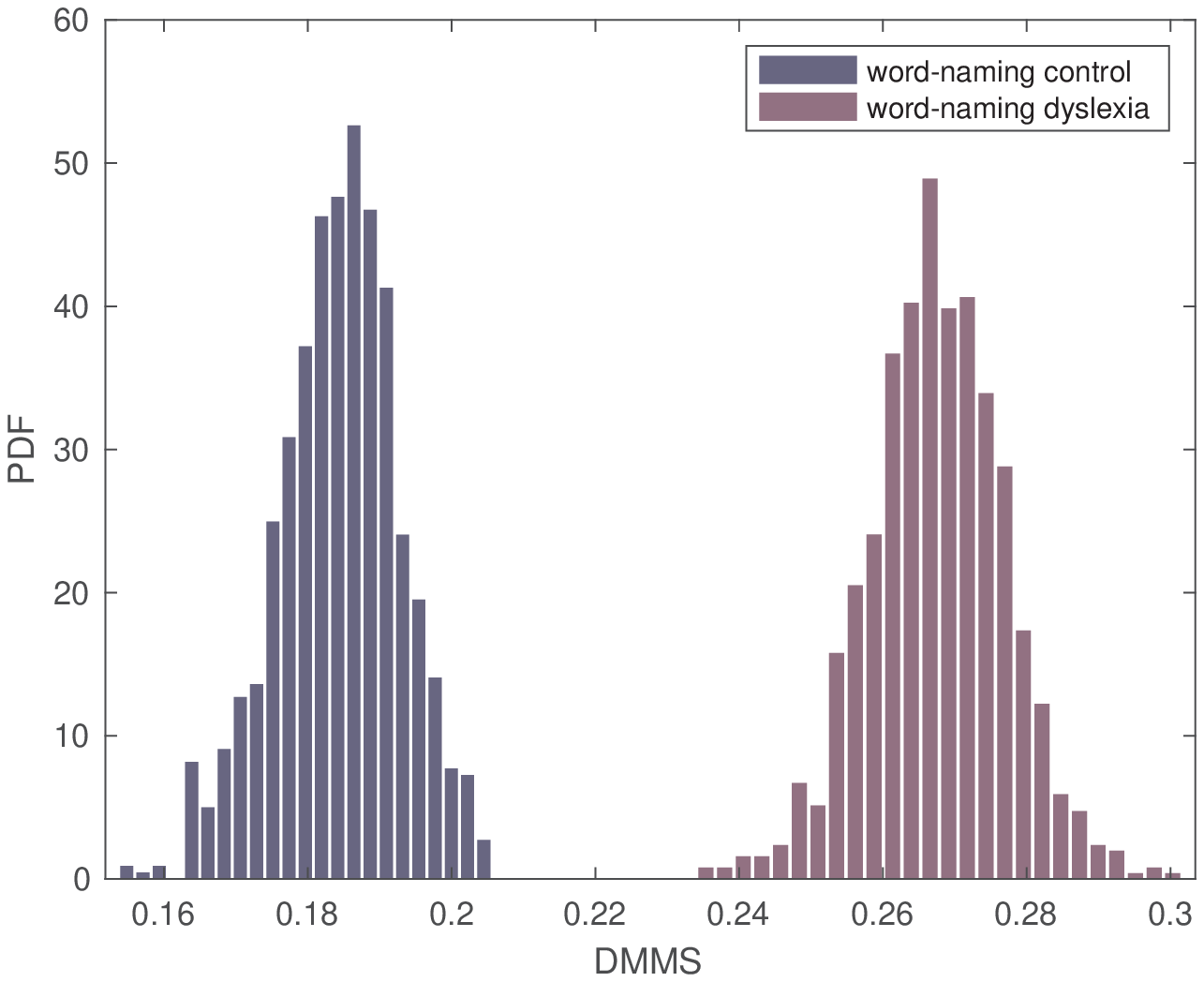}
\includegraphics[width = 0.25 \textwidth]{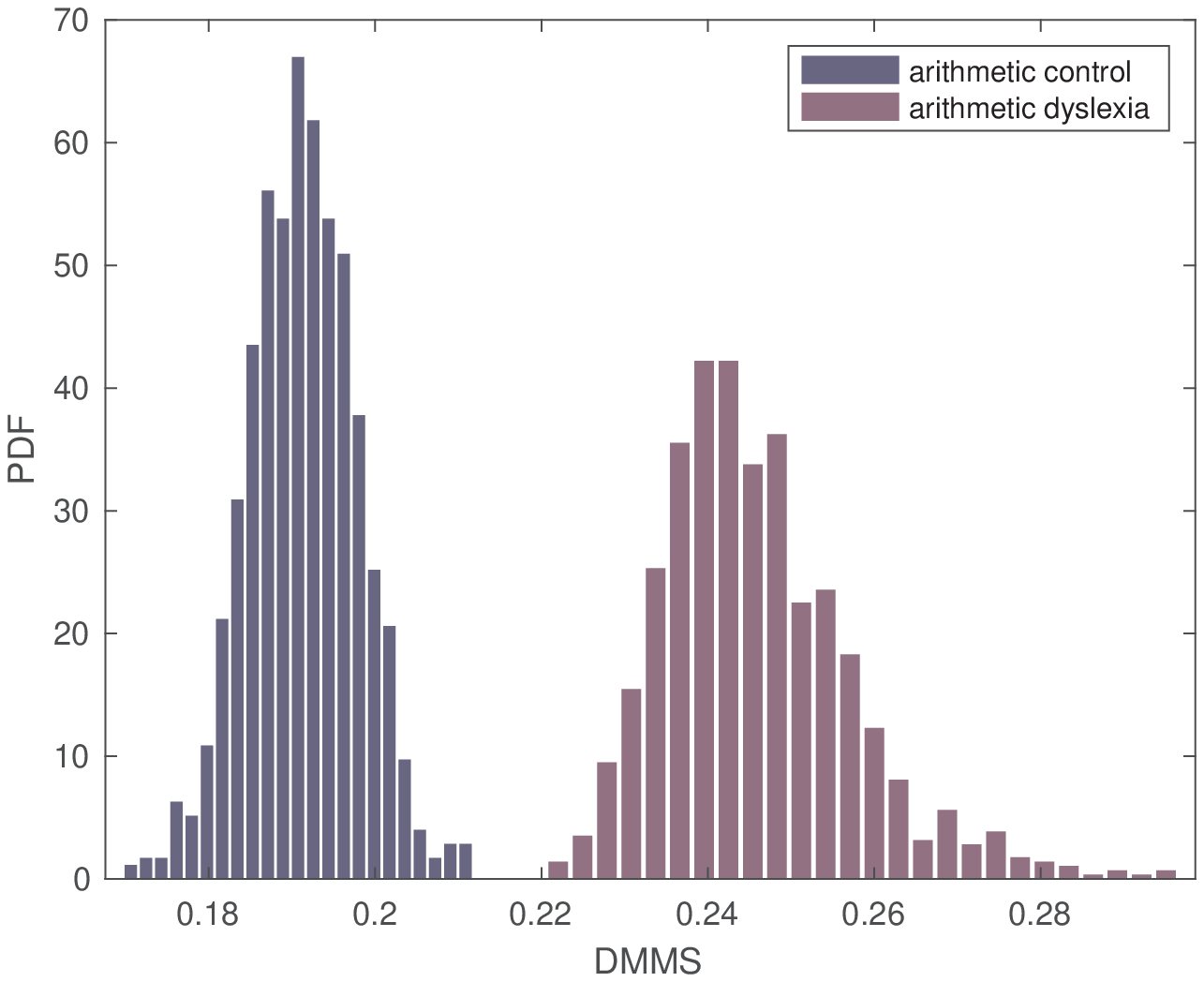} \\
\end{tabular}
\caption{ Bootstrap-obtained distributions of parameters. Columns, from left to right: Flanker, Color Naming, Word Naming, Arithmetic. Rows, from top to bottom: power-law exponent ($\alpha q+1$), Reduced Half Width, Hoover, DMMS.}
\label{figure5}
\end{figure*}

Here we perform bootstrap, as discussed in great detail in \cite{liu2016probability,liu2019modeling}, to establish distributions and CI for the tail exponent $\alpha q +1$ and the variability measures introduced in Section \ref{New} -- rHW, DMMS and Hoover. Fig. \ref{figure5} shows the distributions of these quantities and Table \ref{table1} gives their mean and SD, as well as CI. Sample sizes in bootstrap simulations are very close to those in the study described in \cite{holden2014dyslexic}.

\clearpage

\begin{table}[!htbp]
\centering
\caption{ Mean, standard deviation and 95\% confidence interval for bootstrap results.}
\label{table1}
\begin{subtable}{.5\linewidth}
\caption{Flanker}
\centering
\begin{tabular}{cccrl} 
\hline
 & \multicolumn{4}{c}{Tail Exponent $\alpha q + 1$} \\
\hline
 & Mean & SD & \multicolumn{2}{c}{CI} \\
\hline
C & 4.683  & 0.072 & (4.529, & 4.811) \\
D & 4.603 & 0.072 & (4.430, &  4.731) \\
\hline
 & \multicolumn{4}{c}{Reduced Half Width} \\
\hline
 & Mean & SD & \multicolumn{2}{c}{Range} \\
\hline
C & 0.794 &  0.017 & (0.766, &  0.821) \\
D & 0.722 & 0.023 & (0.684, & 0.760) \\
\hline
 & \multicolumn{4}{c}{Hoover} \\
\hline
 & Mean & SD & \multicolumn{2}{c}{Range} \\
\hline
C & 0.135 & 0.002 & (0.133, & 0.138) \\
D & 0.136 & 0.002 & (0.133, & 0.140) \\
\hline
 & \multicolumn{4}{c}{DMMS} \\
\hline
 & Mean & SD & \multicolumn{2}{c}{Range} \\
\hline
C & 0.205 & 0.006 & (0.195, & 0.215) \\
D & 0.225 & 0.009 & (0.211, & 0.240) \\
\hline
\end{tabular}
\end{subtable}%
\begin{subtable}{.5\linewidth}
\caption{Color }
\centering
\begin{tabular}{cccrl} 
\hline
 & \multicolumn{4}{c}{Tail Exponent  $\alpha q + 1$} \\
\hline
 & Mean & SD & \multicolumn{2}{c}{CI} \\
\hline
C & 4.936  & 0.107 & (4.788, & 5.226) \\
D & 4.335 & 0.098 & (4.198, & 4.570) \\
\hline
 & \multicolumn{4}{c}{Reduced Half Width} \\
\hline
 & Mean & SD & \multicolumn{2}{c}{Range} \\
\hline
C & 0.655 & 0.015 & ( 0.629, & 0.681) \\
D & 0.509 & 0.016 & (0.478, & 0.531) \\
\hline
 & \multicolumn{4}{c}{Hoover} \\
\hline
 & Mean & SD & \multicolumn{2}{c}{Range} \\
\hline
C & 0.135 & 0.002 & (0.132, & 0.1375) \\
D & 0.150 & 0.002 & (0.146, & 0.154) \\
\hline
 & \multicolumn{4}{c}{DMMS} \\
\hline
 & Mean & SD & \multicolumn{2}{c}{Range} \\
\hline
C & 0.182 & 0.008 & (0.169, & 0.196) \\
D & 0.206 & 0.012 & (0.192, & 0.228) \\
\hline
\end{tabular}
\end{subtable}

\begin{subtable}{.5\linewidth}
\caption{Naming }
\centering
\begin{tabular}{cccrl} 
\hline
 & \multicolumn{4}{c}{Tail Exponent  $\alpha q + 1$} \\
\hline
 & Mean & SD & \multicolumn{2}{c}{CI} \\
\hline
C & 7.187  & 0.140 & (6.919, & 7.465) \\
D & 5.217 & 0.077 & (5.063, & 5.367) \\
\hline
 & \multicolumn{4}{c}{Reduced Half Width} \\
\hline
 & Mean & SD & \multicolumn{2}{c}{Range} \\
\hline
C & 0.567 & 0.011 & (0.548, &  0.585) \\
D & 0.424 & 0.011 & (0.405, & 0.443) \\
\hline
 & \multicolumn{4}{c}{Hoover} \\
\hline
 & Mean & SD & \multicolumn{2}{c}{Range} \\
\hline
C & 0.082 & 0.001 & (0.080, & 0.084) \\
D & 0.110 & 0.002 & (0.107, & 0.113) \\
\hline
 & \multicolumn{4}{c}{DMMS} \\
\hline
 & Mean & SD & \multicolumn{2}{c}{Range} \\
\hline
C & 0.184 & 0.008 & (0.169, & 0.198) \\
D & 0.267 & 0.009 & (0.252, & 0.282) \\
\hline
\end{tabular}
\end{subtable}%
\begin{subtable}{.5\linewidth}
\caption{Arithmetic}
\centering
\begin{tabular}{cccrl} 
\hline
 & \multicolumn{4}{c}{Tail Exponent $\alpha q + 1$} \\
\hline
 & Mean & SD & \multicolumn{2}{c}{CI} \\
\hline
C & 4.106  & 0.091 & (3.882, & 4.185) \\
D & 3.232 & 0.090 & (3.072, & 3.423) \\
\hline
 & \multicolumn{4}{c}{Reduced Half Width} \\
\hline
 & Mean & SD & \multicolumn{2}{c}{Range} \\
\hline
C & 0.475 & 0.011 & (0.458, &  0.498) \\
D & 0.295 & 0.015 & (0.269, & 0.320) \\
\hline
 & \multicolumn{4}{c}{Hoover} \\
\hline
 & Mean & SD & \multicolumn{2}{c}{Range} \\
\hline
C & 0.178 & 0.002 & (0.174, & 0.182) \\
D & 0.240 & 0.007 & (0.230, & 0.251) \\
\hline
 & \multicolumn{4}{c}{DMMS} \\
\hline
 & Mean & SD & \multicolumn{2}{c}{Range} \\
\hline
C & 0.191 & 0.006 & (0.181, & 0.209) \\
D & 0.246 & 0.011 & (0.229, & 0.268) \\
\hline
\end{tabular}
\end{subtable}%
\end{table}

\section{Summary \label{Summary}}

We proposed to use Generalized Beta Prime for fitting response-time distributions. It does superior job of fitting relative to previously used distributions and is a steady-state distribution of a simple stochastic model. We used aggregated response times to compare two distinct groups of individuals in a contrast study. In this approach, aggregated group distribution is surmised to reflect on group's underlying cognitive dynamics, while an individual's distribution can be thought of as a random variate of the group distribution. The contrast in distributions was analyzed in terms of overall rescaling, by the ratio of the mean values, and in terms of shape differences. We introduced several shape-related parameters and showed that where the mean values differed the most, the difference in shape parameters was also the greatest.

\clearpage

\appendix

\section{GB2 Fits without Front Cuts\label{Nocut}}
Here, we repeat fitting procedure of Section \ref{GB2fit} but without cutting un-physically small RT, that is to say that we use the full RT data set. 
\begin{table}[!htbp]
\centering
\caption{GB2 fitting parameters and front $\alpha p - 1$ and tail $-(\alpha q + 1)$ exponents.}
\label{GB2wcut}
\begin{tabular}{ccccc}
\hline
\hline
		&	parameters&	   front exponent &    tail exponent &          KS test \\
\hline
flankControl& (          0.7986,           0.3768,		9.6961,           0.4169) &           6.7434 &          -4.6535 &           0.0087 \\
\hline
flankDyslexia& (          0.4992,           0.2755,		12.9841,           0.4530) &           5.4818 &          -4.5767 &           0.0120 \\
\hline
ColorControl& (          0.2282,           0.1971,			18.3427,           0.6298) &           3.1866 &          -4.6163 &           0.0266 \\
\hline
ColorDyslexia& (          0.1690,           0.1170,		26.1264,           0.6825) &           3.4148 &          -4.0580 &           0.0374 \\
\hline
nameControl& (          0.2992,           0.2530,		23.8455,           0.5220) &           6.1337 &          -7.0324 &           0.0265 \\
\hline
nameDyslexia& (          0.1722,           0.1092,		38.3600,           0.6032) &           5.6057 &          -5.1892 &           0.0228 \\
\hline
AirthControl& (          0.6556,           0.4266,			7.0219,           0.9166) &           3.6039 &          -3.9954 &           0.0092 \\
\hline
AirthDyslexia&(          0.3155,           0.1590,		12.7333,           0.9986) &           3.0170 &          -3.0247 &           0.0225 \\
\hline\hline
\end{tabular}
\end{table}

\begin{figure*}[!htbp]
\centering
\begin{tabular}{cc}
\includegraphics[width = 0.35 \textwidth]{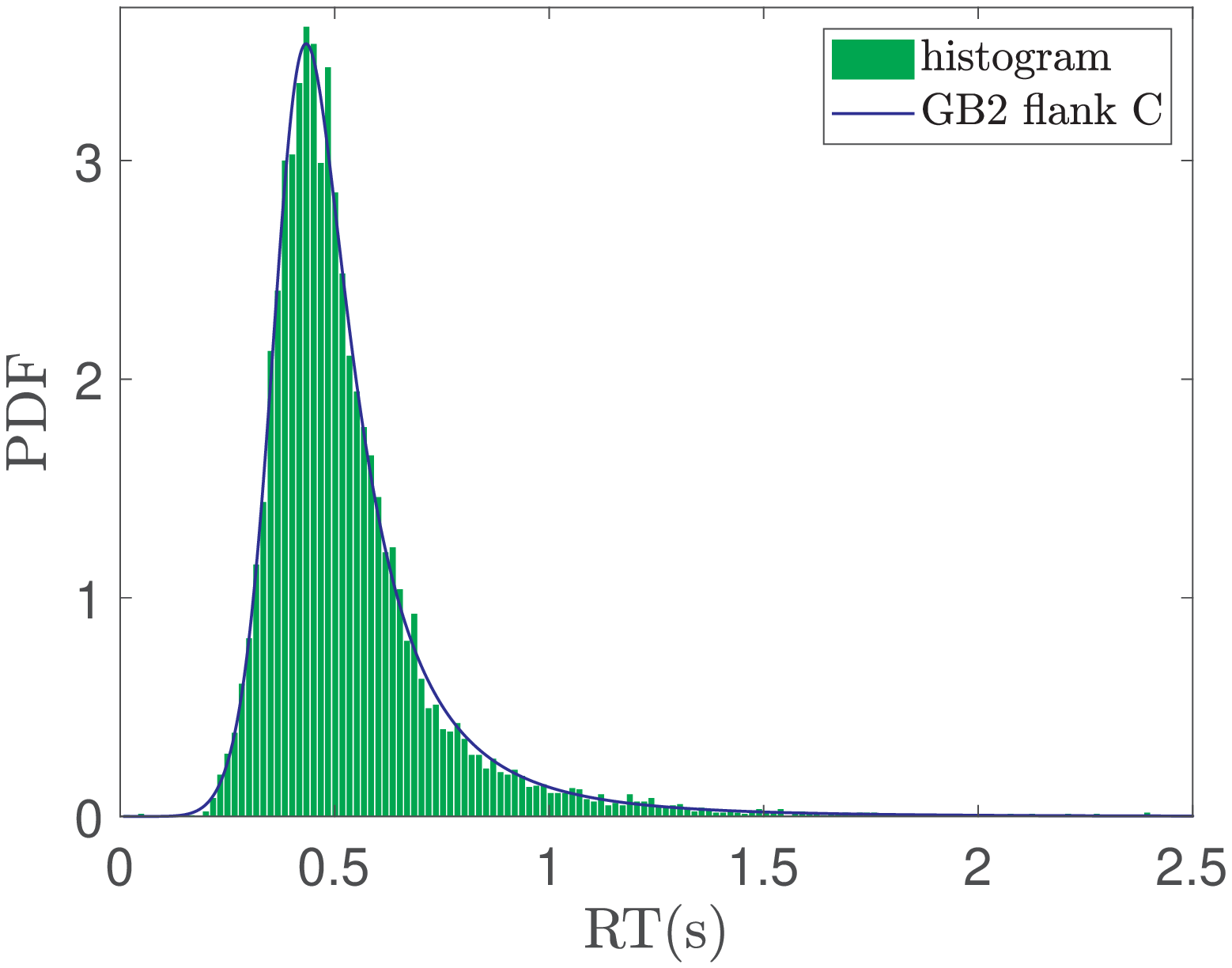}  
\includegraphics[width = 0.35 \textwidth]{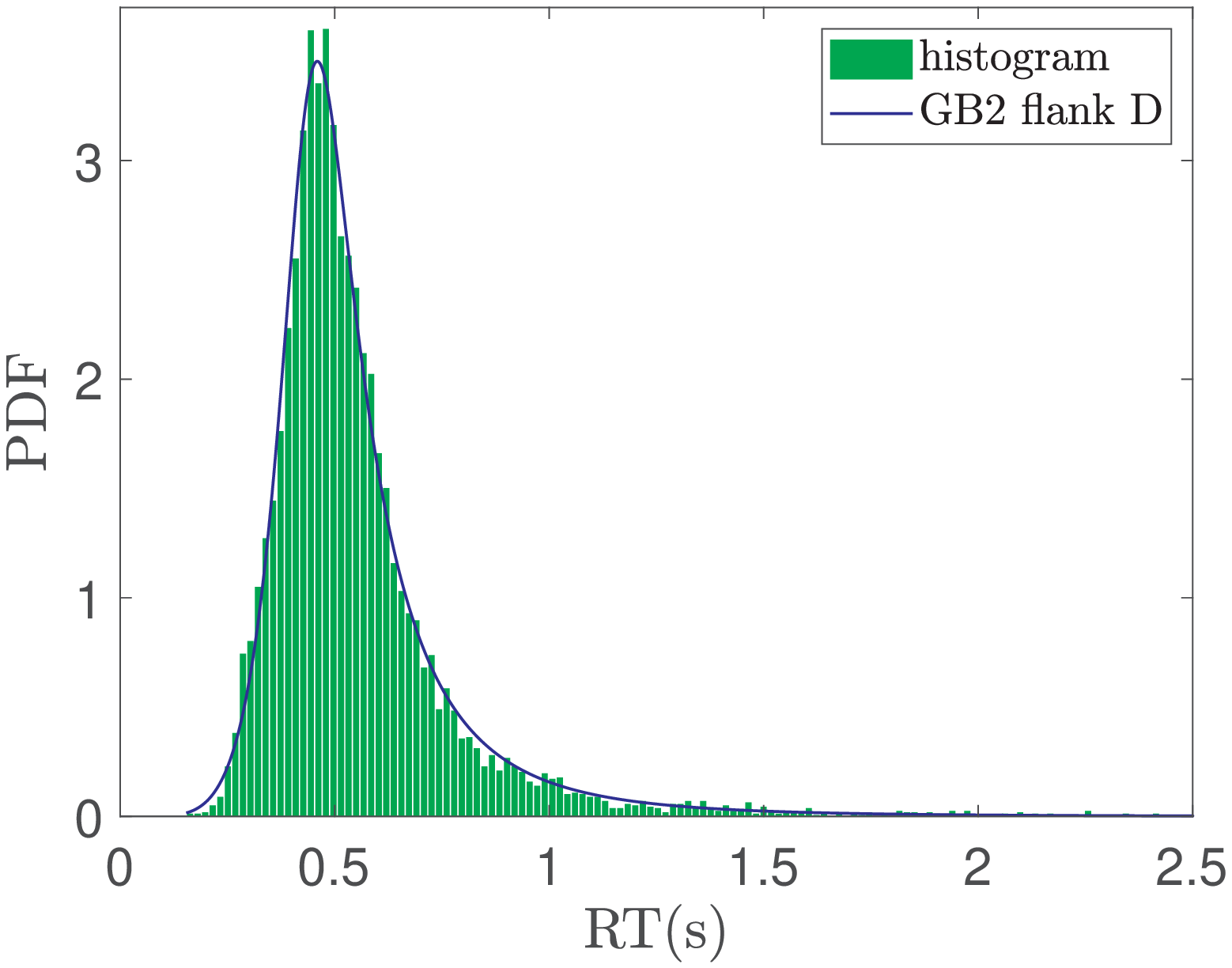}  
\end{tabular}
\caption{Flanker test GB2 fits.}
\label{figure9}
\end{figure*}

\begin{figure*}[!htbp]
\centering
\begin{tabular}{cc}
\includegraphics[width = 0.35 \textwidth]{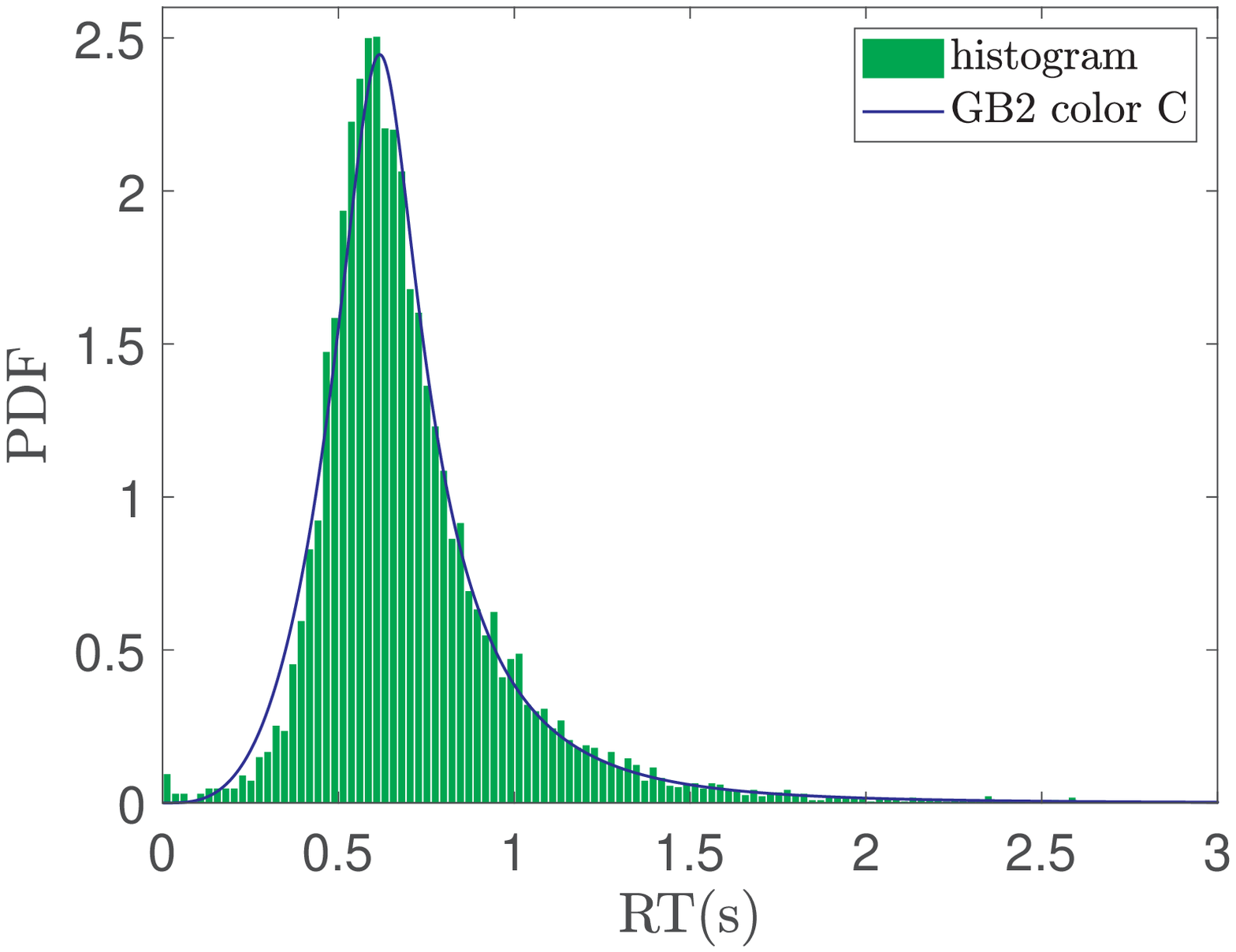}  
\includegraphics[width = 0.35 \textwidth]{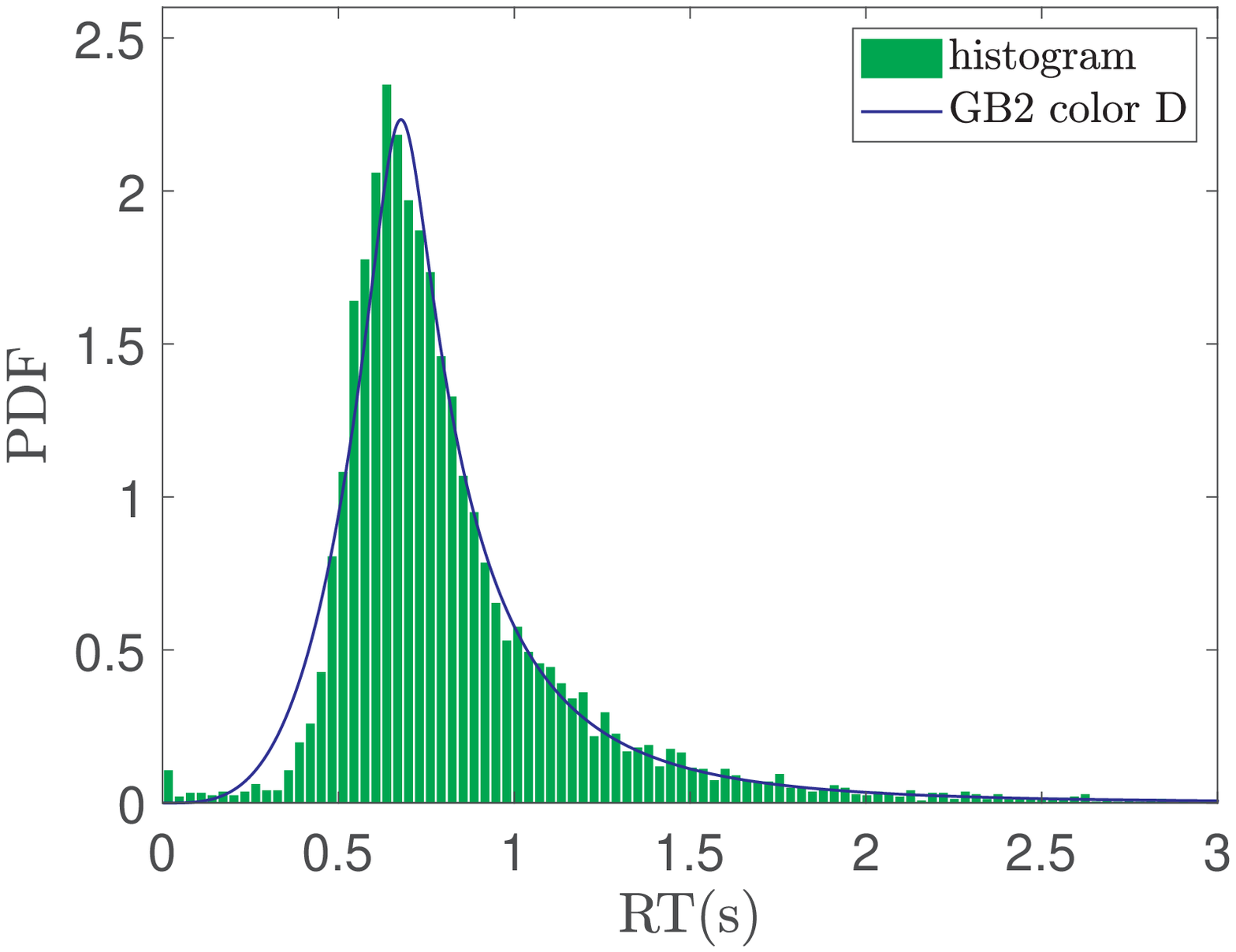}  
\end{tabular}
\caption{Color test GB2 fits.}
\label{figure7}
\end{figure*}

\begin{figure*}[!htbp]
\centering
\begin{tabular}{cc}
\includegraphics[width = 0.35\textwidth]{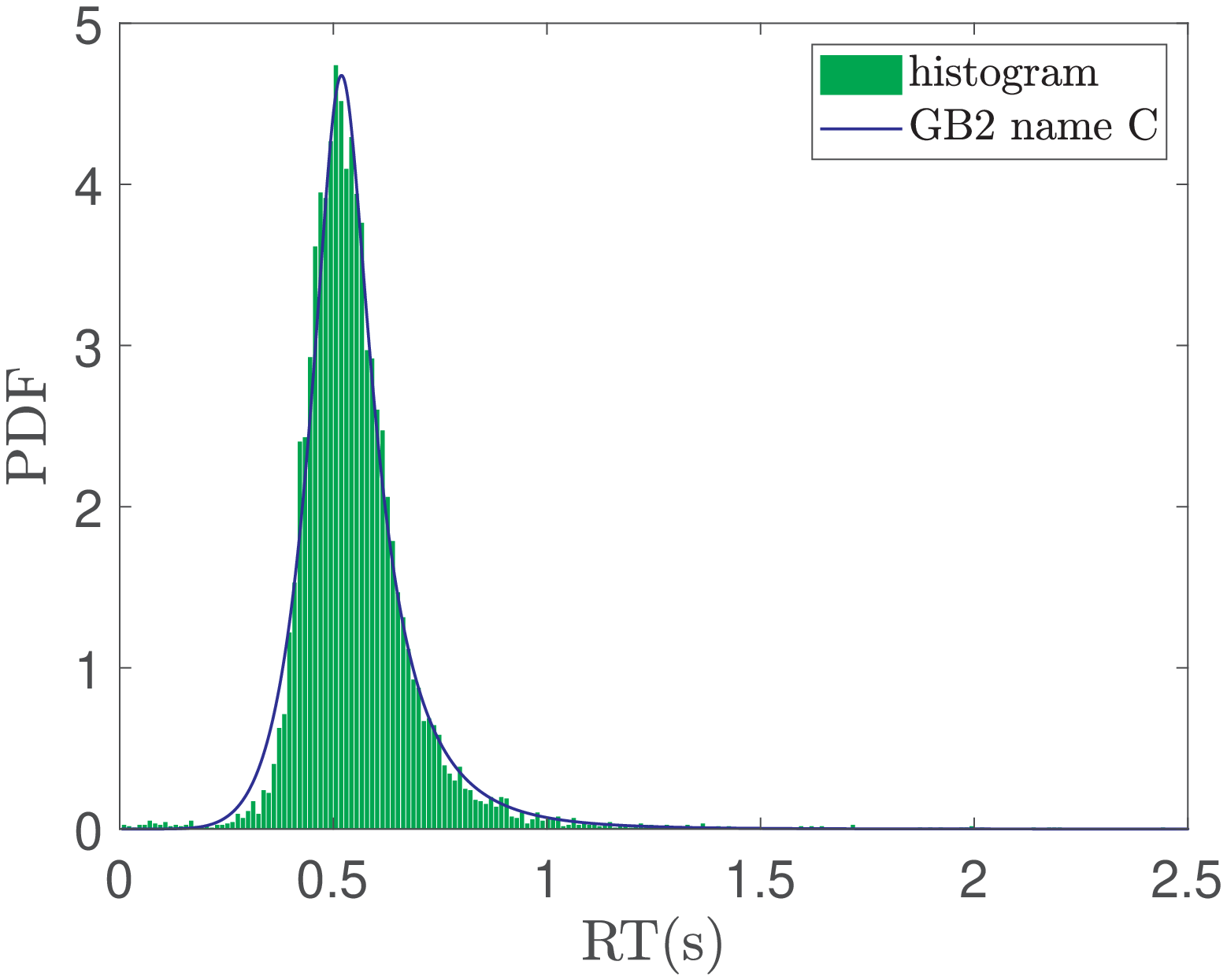}  
\includegraphics[width = 0.35 \textwidth]{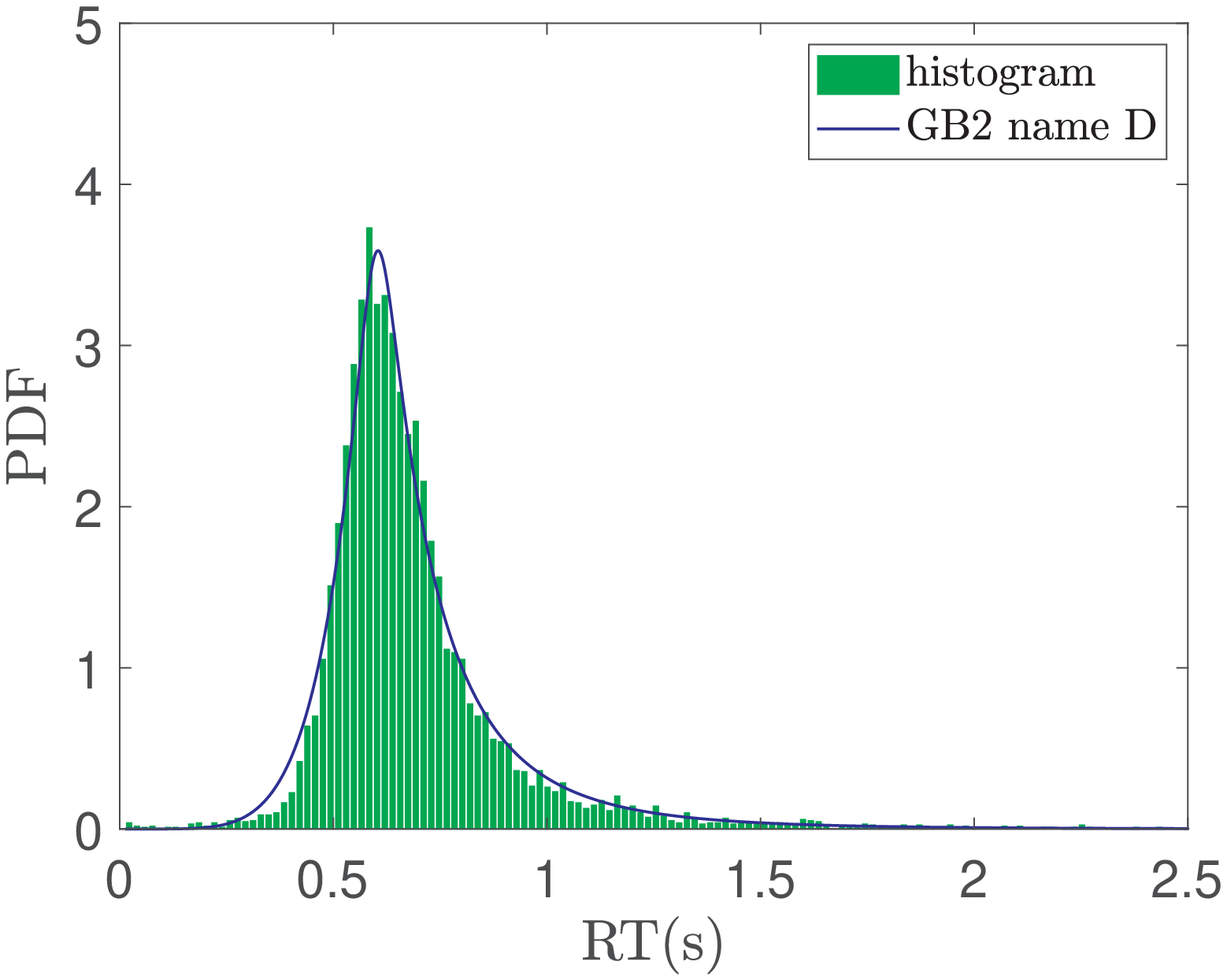}  
\end{tabular}
\caption{Word-Naming test GB2 fits.}
\label{figure8}
\end{figure*}

\begin{figure*}[!htbp]
\centering
\begin{tabular}{cc}
\includegraphics[width = 0.35 \textwidth]{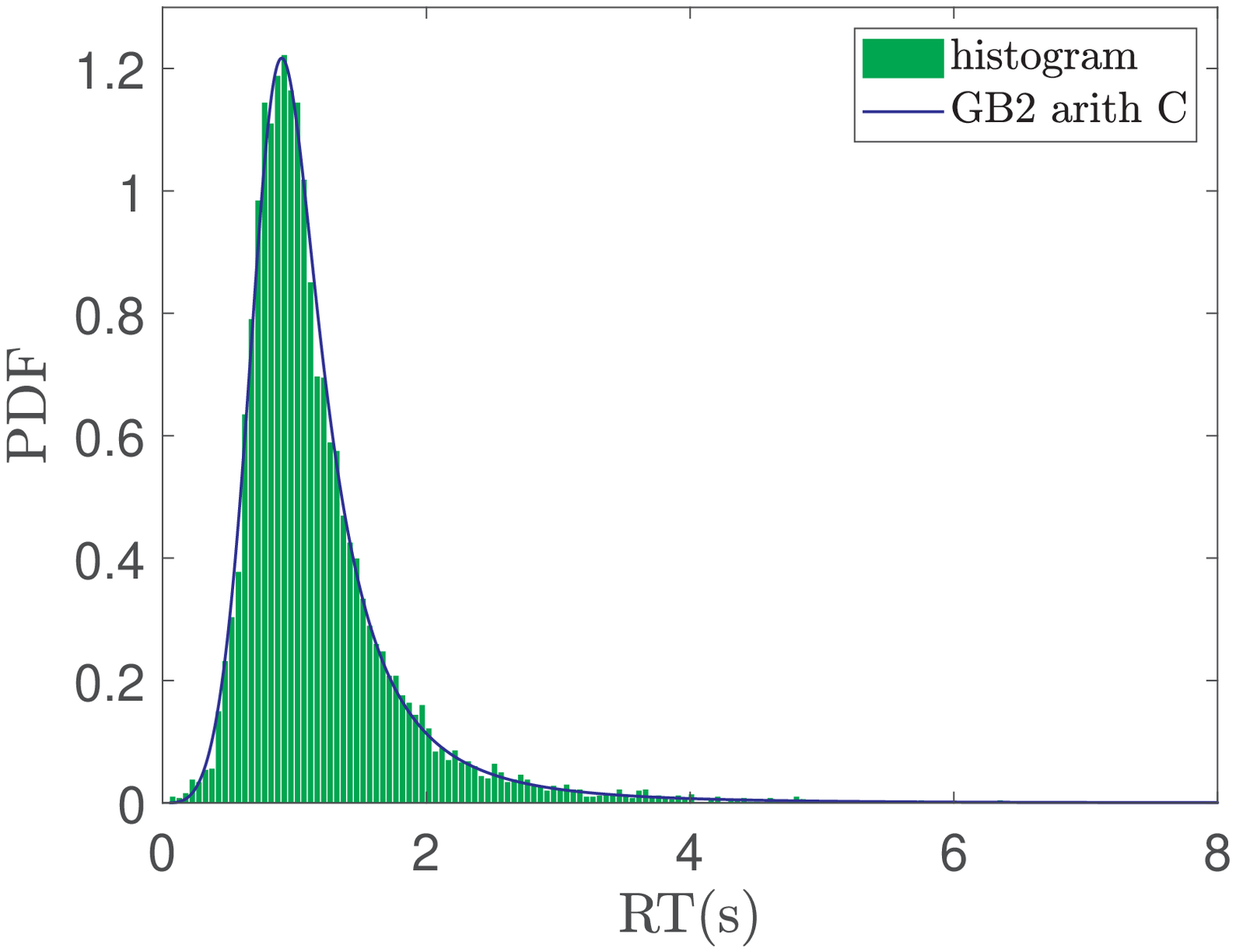}  
\includegraphics[width = 0.35 \textwidth]{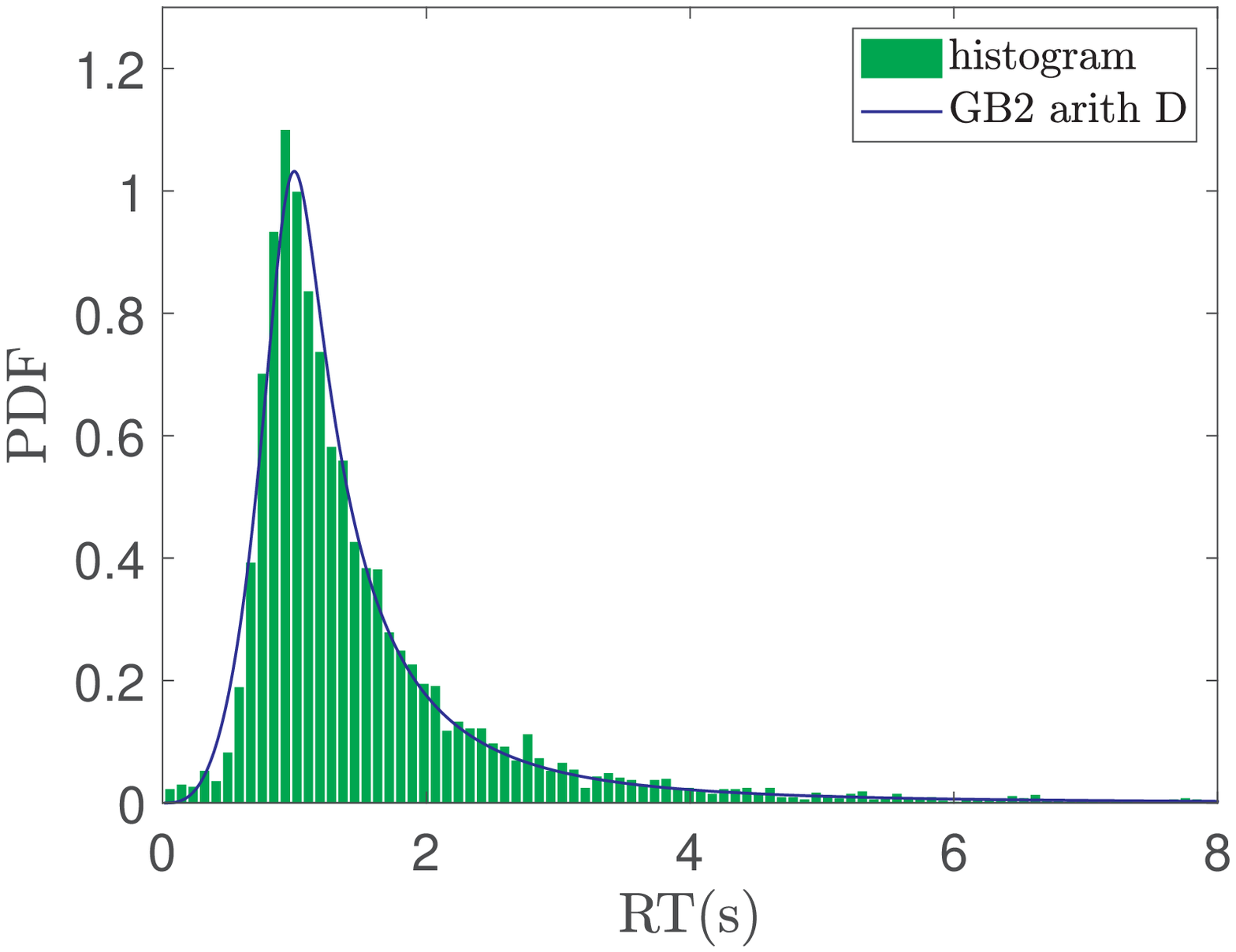}  
\end{tabular}
\caption{Arithmetic test GB2 fits.}
\label{figure6}
\end{figure*}

\clearpage
\bibliography{mybibfile}

\end{document}